\documentclass{jfm}

\usepackage{pdfpages}
\usepackage{subcaption}

\usepackage{graphicx}
\usepackage{newtxtext}
\usepackage{newtxmath}
\usepackage{natbib}
\usepackage{hyperref}
\hypersetup{
    colorlinks = true,
    urlcolor   = blue,
    citecolor  = black,
}

\newcommand{\RomanNumeralCaps}[1]
\linenumbers

\title{Ignition Behind Decaying Shock Waves : Detonation Cells}

\author{Kevin Cheevers\aff{1}
  \corresp{\email{kevin.cheevers@uottawa.ca}},
 \and Matei Radulescu\aff{1}}

\affiliation{\aff{1}University of Ottawa}

\begin{document}
\maketitle

\begin{abstract}
We first address the problem of initiation behind decaying shock waves analytically and numerically. The ignition along a particle path crossing the shock is analysed in terms of its volumetric expansion, evaluated as a function of the shock strength, decay rate, and curvature using the shock change equations. Closed form solutions are derived for 1-step and 2-step chemistry models. The analytical results are found in excellent agreement with numerical simulations using a detailed thermo-chemical model. 
The newly developed model is used to study the ignition behind the lead shock in detonation cells. New experiments are reported for cellular dynamics in a very regular H$_2$-O$_2$-Ar system. It is found that the ignition of approximately half of the gases crossing the shock is quenched. In the experiments, this gas reacts behind the transverse waves. Previous experiments in a very irregular system of methane-oxygen are also analysed. In these experiments, more than 70\% of the gases crossing the shock are quenched due to shock non-steadiness. These gases are shed as non-reacted pockets, which react much slower and burn out due to turbulent diffusion. The main conclusion of the study is that both irregular and regular cellular detonations experience ignition quenching due to shock non-steadiness. 
\end{abstract}

\newpage
\section{Introduction}
\label{ch:Introduction}
\begin{figure}
	\centering
	\begin{subfigure}[c]{0.48\textwidth}
		\includegraphics[width=\linewidth]{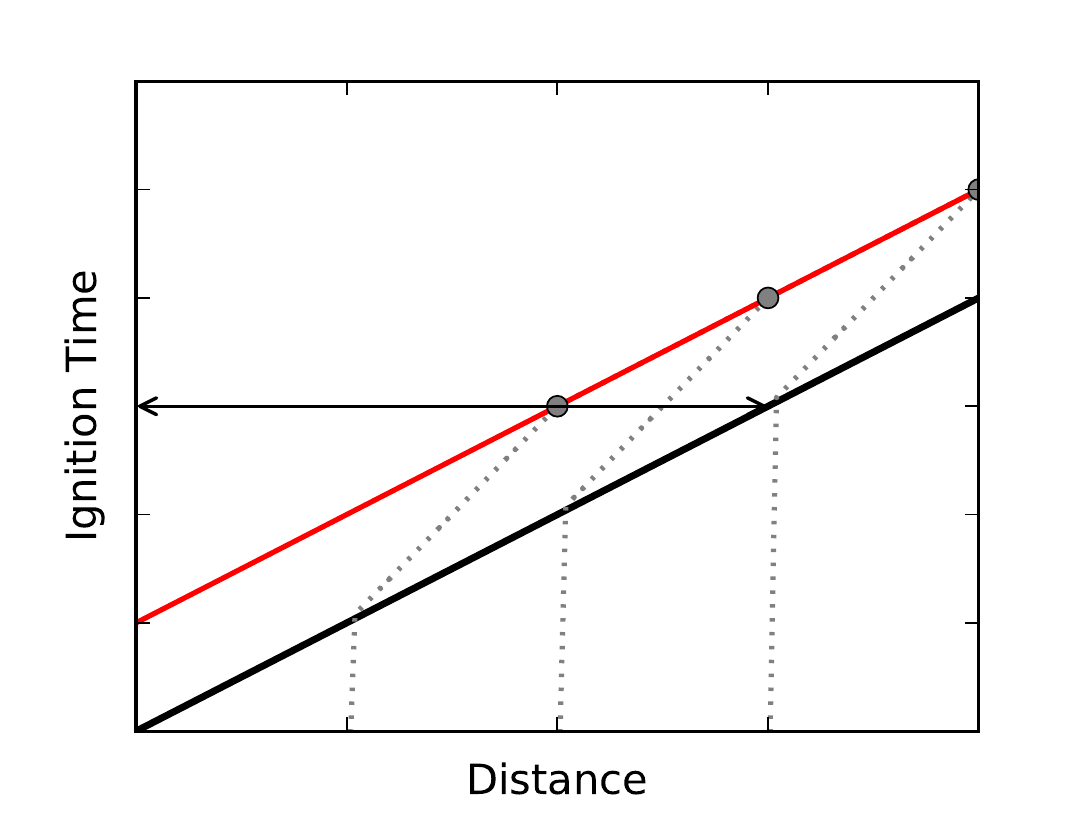} 
		\caption{}
	\end{subfigure}
	\begin{subfigure}[c]{0.48\textwidth}
		\includegraphics[width=\linewidth]{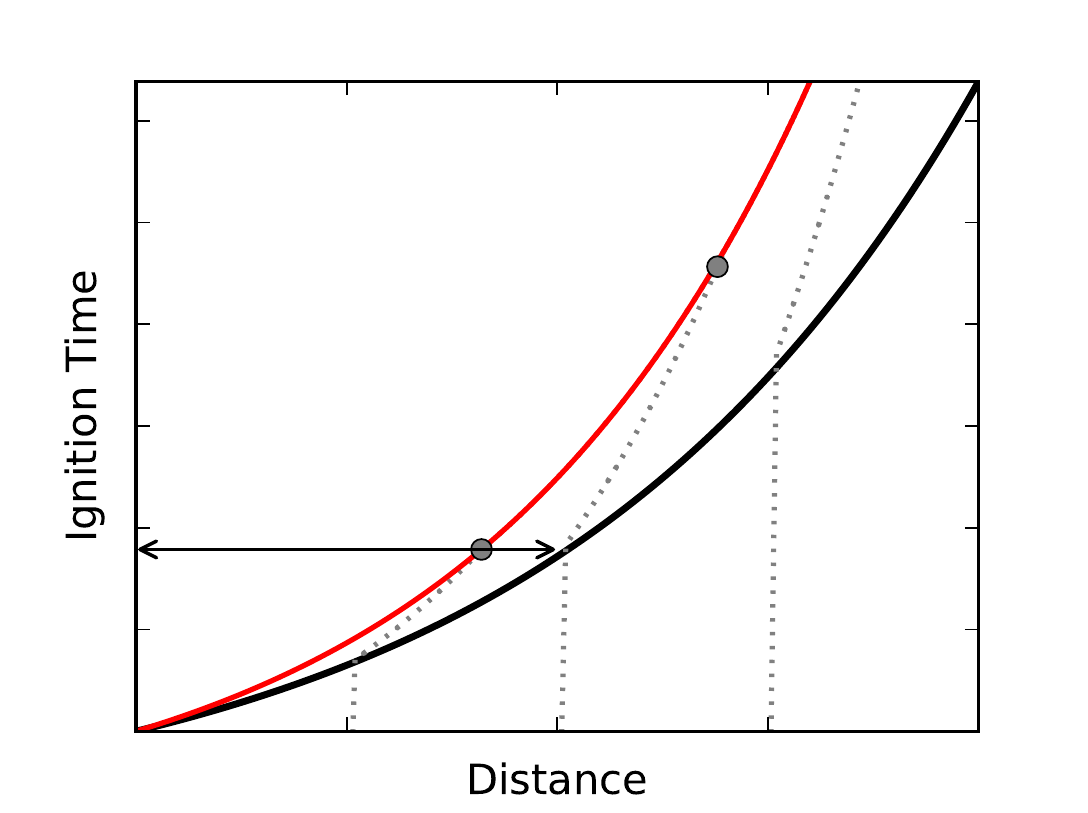} 
		\caption{}
	\end{subfigure}
	\caption{An x-t diagram representing a detonation propagating from left-to-right. a) A steady detonation with a constant velocity, followed by a reaction zone at a fixed distance. All Lagrangian particles crossing the lead shock have the same ignition delay, depending on the strength of the shock. b) An unsteady detonation with a decaying velocity, followed by a reaction zone at an increasing distance. Particles crossing the decaying lead shock have an ignition delay which depends on the strength of the lead shock at the time of crossing.}
	\label{fig:XT_Detonation}
\end{figure}

Detonations are self-sustained combustion waves propagating in a reactive medium at supersonic speeds. The total chemical energy dictates the propagation velocity.  The average speed of propagation observed in experiments is well predicted by Chapman-Jouguet (CJ) theory \citep{LeeBook2008}. Real detonations have a multidimensional cellular structure which emerges due to the feedback effect caused by the sensitivity of the induction time to changes in temperature behind the lead shock, and the dependance of the strength of the lead shock to changes in the post-shock rate of energy release. This in turn makes it difficult to measure the induction time of a reactive lagrangian particle which crosses a decaying shock at a known time. Due to the decaying nature of the lead shock throughout detonation cells, a direct measurement of the induction zone is not representative of the induction time, as shown in Figure \ref{fig:XT_Detonation}. The variation of the shock speed can be quite large during a cell cycle, ranging from approximately $1.8 D_{CJ}$ at the beginning of the cell cycle, to approximately $0.6 D_{CJ}$ at the end of the cell \citep{Strehlow&Crooker}, where $D_{CJ}$ is the propagation velocity determined from CJ theory.

\begin{figure}
	\centering
	\hfill
	\begin{subfigure}[c]{0.4\textwidth}
		\includegraphics[width=\linewidth]{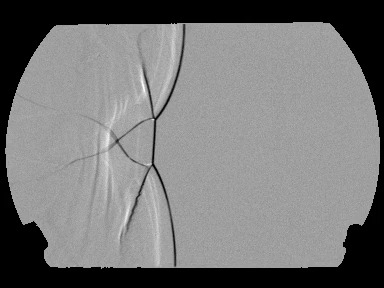}
		\caption{}
	\end{subfigure}
	\hfill
	\begin{subfigure}[c]{0.4\textwidth}
		\includegraphics[width=\linewidth]{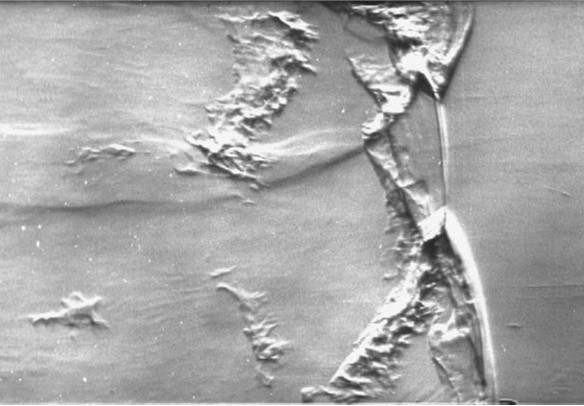} 
		\caption{}
	\end{subfigure}
	\hfill
	\hfill
	\caption{Schlieren visualization of detonations propagating from left-to-right through a) a stable mixture of 2H$_2$+O$_2$+7Ar \citep{Xiao2020}, and b) a highly unstable mixture of CH$_4$ + 2O$_2$, with visible turbulent mixing pockets of unreacted flow in the post-shock region \citep{Radulescu2007}.}
	\label{fig:CellularInstability}
\end{figure}

\citet{Subbotin1975} was the first to identify two types of detonation cellular structures, the difference between the two lying in the mechanism of combustion and the regularity of detonation cell structure. Weakly unstable detonations have very regular structures, with the most regular structure registered in the hydrogen-oxygen-argon system, as illustrated in Figure \ref{fig:CellularInstability} a).  In contrast, highly unstable detonations have irregular structures, an example being methane-oxygen detonations showed in Figure \ref{fig:CellularInstability} b). The main feature of weakly unstable detonations is the prompt termination of the induction zone close the lead shock and transverse waves.  In contrast, the main features in highly unstable detonation cells are the same Mach stem, incident shock, transverse waves, and reaction waves as in the stable detonations, but one also sees pockets of unburnt gas further downstream, which burn via diffusive phenomena (wrinkled flames) \citep{Xiao2020}. The increased delays in combustion for these non-reacted pockets is due mainly to the higher sensitivity of the induction time to shock temperature in these systems.  

\citet{RadulescuPhDThesis} previously categorized the instability (and conversely the regularity) of a detonable mixture as a function of its $\chi$ parameter . This is found to be a function of the mixture's sensitivity, and the ratio of the induction time to explosion time, 
\begin{equation}
\chi = \frac{Ea}{RT}\frac{t_i}{t_r}.
\end{equation}
Mixtures with small $\chi$ parameters, such as 2H$_2$ + O$_2$ + 7Ar, are weakly unstable and their cellular structure are similar to that shown in Figure \ref{fig:CellularInstability} a). Higher $\chi$ parameters are associated with unstable mixtures, such as CH$_4$ + 2O$_2$, and their cellular structure resembles that shown in Figure \ref{fig:CellularInstability} b). Distinguishing between mixtures is important when considering that the secondary initiation mechanisms are different for each structure. 
The two fuel mixtures used in the current study were 2H$_2$ + O$_2$ + 7Ar and CH$_4$ + 2O$_2$, with $\chi$ parameters of 4.7 and 11000, representing a weakly unstable mixture which is insensitive to small changes in temperature and a very unstable mixture which is extremely sensitive to small changes in temperature. These mixtures are chosen because of the extremes they represent, allowing for the observation of generalities across the entire range of detonable mixtures.

Modelling and predicting the effective energy release rate behind the front of cellular detonations is very important for predicting the dynamics of detonations in practical problems such as detonation arrestors, detonation engines, and cloud ignition. The prediction of this rate of energy release behind the lead shock in cellular detonations remains very challenging. With current numerical simulations, it is currently not feasible to perform well-resolved numerical simulations of detonations for macroscopic applications due to the large range of scales present. The smallest scale required for such simulations is on the order of $10^{-7}$ m to properly resolve the reaction zone structure, the thin reaction zones and its interaction with turbulence at the Kolmogorov scale \citep{Powers2006}. In contrast, the largest scales in applications are on the order of the device size, and approximately 10 to 1000 larger than the characteristic cell dimension. This \textit{outer} scale is on the order of meters. To solve these problems, one requires a macroscopic model which can accurately predict the energy release occurring behind the lead shock in detonations.  Recognizing the non-stationarity of the lead shock at scales smaller than the cell size, the first step would thus be studying the effect of the unsteadiness of the shock on the reaction zone structure, such that one can predict the distribution of ignition delays in the reaction zone structure of detonations. 

The problem of ignition within the structure of an unsteady cellular detonation can be posed as a generic problem of ignition behind a decaying shock wave.  If the shock dynamics (its curvature, speed and acceleration) are locally known \textit{a priori}, the ignition dynamics can be determined. This problem consists of a Lagrangian particle that crosses a shock whose strength decays with time, after which the particle undergoes an expansion. This particle expansion can be visualized as a zero-dimensional combustor problem whose volume change in time is described in terms of the dynamics of the shock. To solve these problems and thus be able to predict the ignition of a particle behind the shock, one needs to know the relation between the shock dynamics and the time history of the volume change of the Lagrangian particle after it crosses the shock.

\citet{Lundstorm&Oppenheim1969} first studied this problem to address the ignition in cellular detonations in mixtures of hydrogen-oxygen diluted with nitrogen , which have a fair regularity and lie between the two extremes discussed above. The dynamics of the lead shock were approximated as those of a Taylor-Sedov blast wave. Using blast wave theory, they inferred the rate of particle expansion along particle paths and used this to evaluate the particle's ignition delay under the influence of a decaying lead shock. Using ignition kinetics derived from shock tube experiments, they found that ignition behind the lead shock is quenched within the first half of the cellular cycle. The gas particles that cross the lead shock after this quenching point are ignited through other secondary mechanisms, such as heating by transverse shock waves.

More recently, \citet{eckett2000role} developed an approximate solution by reformulating the reactive inviscid Euler equations in the vicinity of the shock wave to treat the evolution of ignition . Their local analysis of the competing terms for the rate of energy release along particle paths revealed the possibility of ignition quenching.   This critical ignition was associated with a critical decay rate of the lead shock. For decay rates higher than this critical decay rate, the post-shock reaction is quenched reminiscent to the quenching in the model of Lundstrom and Oppenheim.  At around the same time as Eckett's work, \citet{vidal1999} also independently analyzed this same problem and arrived at similar conclusions.  

\citet{Austin2003} further used Eckett's model to study the decoupling of the decaying lead shock and the subsequent reaction wave in numerical simulations of idealized cellular detonations.  She analysed the numerical simulations performed by Gamezo and co-workers of a cellular detonation propagating through a 2-dimensional channel \citet{Gamezo1999} described by a one-step Arrehnius reaction. The activation energy of the gas was increased between simulations to increase the instability of the detonation. Austin used the time evolution of the lead shock velocity along the centreline of the tube to calculate the shock decay. Comparing the critical decay rate calculated by applying Eckett's model to data measured from results obtained from the numerical simulation, local decoupling of the reaction wave and lead shock is found for unstable detonations with high activation energy. The decay rate of the lead shock at the intermediate activation energy is not comparable to the critical decay until the very end of the cycle, thus no local decoupling was predicted. 

The problem of ignition behind decaying shock waves has also been addressed by \citet{Kiyanda2013}, who performed photographic studies of cellular detonations of methane-oxygen , which have a very irregular structure. Similarly to Lundstrom and Oppenheim, they used blast wave theory for inert gases to determine the evolution of temperature as a function of time along a particle path. The evolution of temperature with time was then used as a forcing function during the integration of a homogeneous reactor modelling a Lagrangian particle. This phenomenological model found quenching halfway through the cellular cycle.  The experiments suggested that turbulent mixing of unreacted flow pockets with products of combustion were the the secondary initiation mechanism.

More recently, an alternative solution to the problem of shock induced ignition behind non-steady shocks has been proposed by \citet{Radulescu&Maxwell2010}. They formulated their model using the energy and species evolution equations, with the rate of change of volume of a particle as a function of time appearing explicitly in their formulation.  The volumetric forcing function can be predicted using arbitrary shock dynamics in the Newtonian limit or by using a local analysis at the shock using the shock change equations \citep{Radulescu2020_ShockChange}. In their analysis, using one-step chemistry, critical ignition was also found to depend on the volumetric expansion rate of the Lagrangian particles in the post-shock state.

Previous research have thus showed that expansion induced quenching was present in unstable detonations, while the situation was not clear for weakly unstable ones, idealized simulations showing that was not the case. The models differed between the authors. The present study attempts to unify the theoretical treatment for ignition behind decaying shocks. It addresses this problem of ignition behind decaying shocks in the context of cellular detonation waves, at the two extremes of cellular regularity.  

This present work will study the results of shock tube experiments reported for weakly unstable detonations in a mixture of argon-diluted hydrogen-oxygen, since it is the system with the most regular structure known to date.  In contrast, analysis of highly unstable detonations in methane-oxygen mixtures reported in our group previously are also analysed to further compare with the findings of Kiyanda and Higgins. In these systems, high speed photographic sequences of the experiments are used to determine the dynamics of the lead shock.  The rate of volumetric expansion along a particle path is inferred from these measurements using the shock change equations. The knowledge of these source terms allows the integration of the energy and species equations in order to study ignition. The problem is addressed numerically for realistic multi-specie kinetic models and theoretically for one and two step reaction formulations. These simplifications allow analytical treatments using asymptotic methods.
\section{Shock Dynamics to Predict $\frac{1}{\rho}\frac{D\rho}{Dt}$}
\label{ch:ShockDynamics}
The analysis of the shock dynamics comprises linking the evolution of the lead shock in time to the evolution of the post-shock particle expansion rate. This is done by using a shock change equation, derived from the conservation equations of mass, momentum, and energy \citep{Radulescu2020_ShockChange}. 

If one considers the state immediately behind the shock in the induction zone, the thermicity $\dot{\sigma}$ is found to be negligible, the shock change equation can be written as
\begin{equation}
\frac{1}{\rho}\frac{D\rho}{Dt} = \frac{u\kappa(1-\eta)+\frac{1}{\rho c^2}\left(\frac{dp}{dt}\right)_S\left(1+\rho_0 (D-u_0)\left(\frac{du}{dp}\right)_H\right)}{\eta},
\label{eq:ShockChange}
\end{equation}
which relates the rate of particle expansion to the variables describing the shock dynamics, such as the shock speed $D$, its acceleration through the rate of change of the post shock pressure $\left(\frac{dp}{dt}\right)_S$, and its curvature $\kappa$. In this equation, $\eta$ is the sonicity 
\begin{equation}
\eta = (1-M)^2,
\end{equation}
in which M is the Mach number of the flow behind the shock relative to the shock wave. All variables without subscripts refer to properties evaluated at the post-shock state, which are dependent on a single variable measureing the shock strength through the shock jump relations and the initial state density $\rho_0$  and the particle velocity $u_0$. The term $\left(\frac{du}{dp}\right)_H$ represents the change of post-shock particle speed with the post shock pressure, which is a property of the mixture measured along the shock Hugoniot curve. The shock change equation can also be written in terms of the specific volume $v = \frac{1}{\rho}$, or
\begin{equation}
\frac{1}{\rho} \frac{D\rho}{Dt} = -\frac{1}{v} \frac{Dv}{Dt}.
\end{equation}
Equation (\ref{eq:ShockChange}) can be further simplified by the assumption of a perfect gas, with $\gamma$ the ratio of specific heats and $M_s$ the propagation Mach number of the shock with respect to the flow ahead of the shock, i.e., $M_s=(D-u_0)/c_0$. By subsequently assuming a strong shock and zero flow speed ahead of the shock, the shock change equation reduces to 
\begin{equation}
\frac{1}{\rho}\frac{D\rho}{Dt} = \frac{6}{\gamma+ 1}\frac{\dot{D}}{D} + 2D\kappa\frac{\gamma-1}{\left(\gamma+1\right)^2},
\label{eq:StrongShockChange}
\end{equation}
in which $\dot{D}$ is the acceleration of the shock. This form of the shock change equation separates the two terms contributing to the gas volumetric expansion behind the shock: the shock unsteadiness and lateral strain rate. The working assumption is made that the volumetric expansion rate evaluated immediately behind the shock remains constant throughout the induction zone, which will permit modelling the ignition process.

\section{Ignition Along the Particle Path}
\label{ch:IgnitionDescription}
Consider a particle of gas at some initial state which eventually crosses a shock at time $t_s$. This inert shock will increase the pressure, temperature, and density within this particle of gas in accordance with the shock jump relations.  From the moment the particle crosses the lead shock, this problem becomes that of a particle of gas subjected to a volumetric expansion in time, with the expansion exerted on the particle found using the shock change equation mentioned in the previous section. What remains in order to solve this problem is to integrate the state within this particle of gas in time. If the time history of density (or volume) is known for a fluid particle crossing the shock wave, the process of ignition of a fluid particle is described by the chemical kinetic laws describing the production or depletion of the different species at the prescribed density, as well as the energy budget.

The characteristic rate of expansion $\alpha$ 
\begin{equation}
\alpha = \frac{1}{v}\frac{Dv}{Dt} = -\frac{1}{\rho}\frac{D\rho}{Dt}, \label{eq:ExpansionRate}
\end{equation}
experienced by the particle is the value it has acquired when crossing the shock wave, given by the shock dynamics via the shock change equation.  In the foregoing, we assume each particle retains this characteristic expansion rate.  With this approximation, the exact energy equation for diffusionless flow, namely, 
\begin{equation}
c_v \frac{DT}{Dt} = - \sum_{i=1}^N e_i {\frac{DY_i}{Dt}}+\frac{p}{\rho^2}\frac{D\rho}{Dt} 
\label{eq:CanteraEnergy}
\end{equation}
simplifies to an ordinary differential equation:
\begin{align}
c_v \frac{dT}{dt} &=- \sum_{i=1}^N e_i {\frac{dY_i}{dt}} -\alpha \frac{p}{\rho}.   \label{eq:energyode}
\end{align}
The relations describing the evolution of chemical compositions along the particle path for diffusionless flow, namely,
\begin{align}
\rho \frac{DY_i}{Dt} &= \dot{\omega}_i W_i,
\end{align}
are also ODE's, which can be re-written for clarity as 
\begin{equation}
\rho \frac{dY_i}{dt} = \dot{\omega}_i W_i. \label{eq:speciesode}
\end{equation}
In these equations, $\dot{\omega}_i$ is the molar production rate of $i$ per unit time, per unit volume.  It can be evaluated from the law of mass action from chemical kinetics, given the rate of individual reactions.

The problem to solve is thus described by the initial value problem given by the two ODE's \eqref{eq:energyode} and  \eqref{eq:speciesode}, 
with initial conditions $Y_i = Y_{i,0}$, and $T = T_s$ evaluated immediately behind the shock.

\subsection{Multiple Species Real Gas Calculation}
\label{ch:CanteraDescription}
The ignition problem described in the previous section can be solved numerically for realistic thermo-chemical data of reactive mixtures of interest.  To this end, we use Cantera, which is a suite of tools developed to solve numerical problems involving chemically reacting flows \citep{Cantera}. This package handles the integration of the species and energy conservation ODEs described in section \ref{ch:IgnitionDescription}. The problem of a particle undergoing a volumetric expansion in time can be modelled in Cantera using the zero-dimensional homogeneous reactor package. This models a Lagrangian fluid element which has a homogeneous state and species composition at every point in time, and no gradients in any spatial dimension within the control system. The initial state of the particle upon crossing the lead shock is calculated using the \textit{fr\_postshock} method available in the Shock and Detonation Toolbox developed by \citet{SDToolbox}. The chemical model used for the integration is the San Diego 2016 model developed by \citet{SanDiego}, containing a total of 57 species.

\begin{figure}
	\centering
	\begin{subfigure}[c]{0.48\textwidth}
		\includegraphics[width=\linewidth]{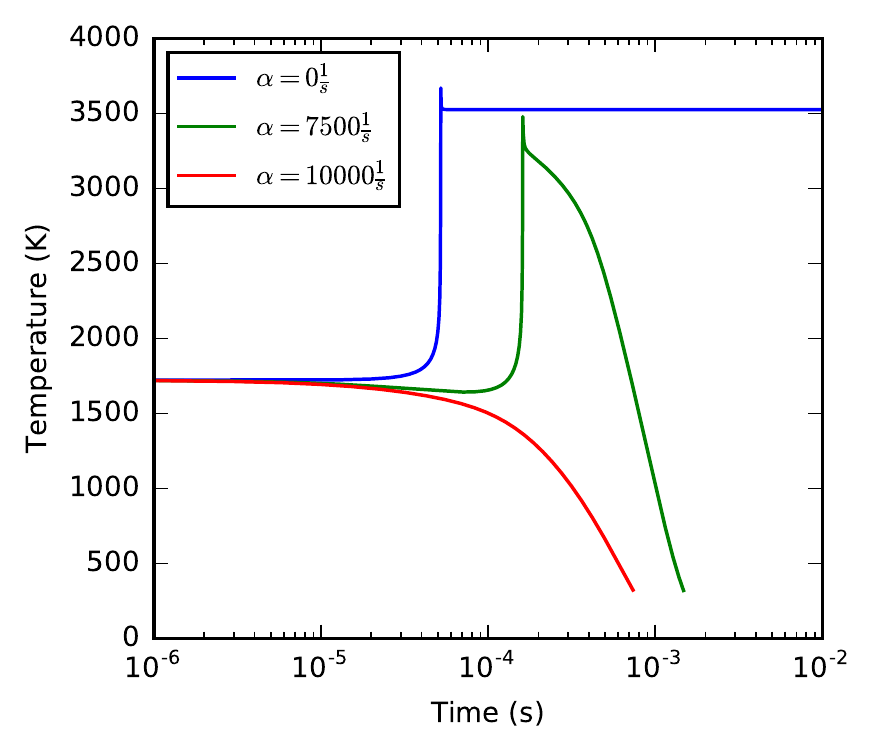} 
	\caption{}
	\end{subfigure}
	\begin{subfigure}[c]{0.48\textwidth}
		\includegraphics[width=\linewidth]{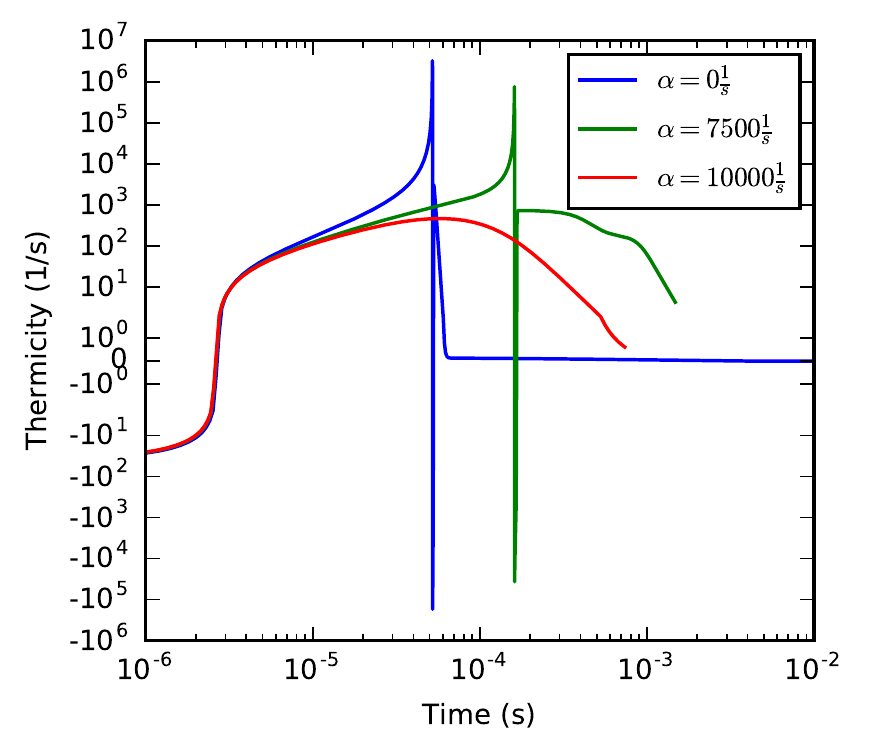} 
		\caption{}
	\end{subfigure}
	\caption{Results of the integration of a shock wave propagating through a quiescent mixture of stoichiometric methane-oxygen, with an initial pressure and temperature of 4.1 kPa and 293 K, for increasing expansion rates $\alpha = -\frac{1}{v}\frac{Dv}{Dt}$. a) temperature vs time, and b) thermicity vs time.}
	\label{fig:CJPostShock}
\end{figure}

As an example, consider a shock wave propagating at the CJ velocity into a mixture of stoichiometric methane-oxygen at an initial state of 4.1 kPa and 293 K. The post-shock state is subjected to increasing expansion. The aforementioned equations are integrated to visualize the evolution of the chemical processes behind the shock. The evolution in time of the post-shock temperature and rate of heat release are shown in Figure \ref{fig:CJPostShock}. Three curves are included on each graph, representing a shock which propagates at a constant CJ velocity without any decay, and two shocks, which propagate at CJ velocity with increasing decays. The shock propagating with no expansion has a very noticeable jump in temperature coinciding with a spike in the rate of heat release, which occurs at the ignition time following an induction zone with relatively little heat release or rise in temperature. The second curve is that of a post-shock state with increased post-shock expansion. This curve sees an increased ignition delay due to the expansion felt by the shocked gas. In addition to the increased ignition delay, one notices that the peak temperature and rate of heat release reached are lower than those of the gas without any expansion. Due to the expansion in this curve, one notices that the post-combustion temperature to steadily decrease once the heat release becomes negligible due to expansion cooling. If one increases the post-shock expansion even further, the expansion of the gas will dominate the post-shock chemical kinetics such that the gas will not react, as the cooling overcomes the effect of post-shock heat release. In this case, no noticeable increase in temperature occurs, and the rate of heat release remains smaller by a few orders of magnitude than those associated with combustion. One can deduct from these figures that a critical expansion rate exists, with smaller expansion allowing for ignition behind the lead shock and higher expansion leading to the quenching of the post-shock combustion.

\begin{figure}
	\centering
	\begin{subfigure}[c]{0.48\textwidth}
		\includegraphics[width=\linewidth]{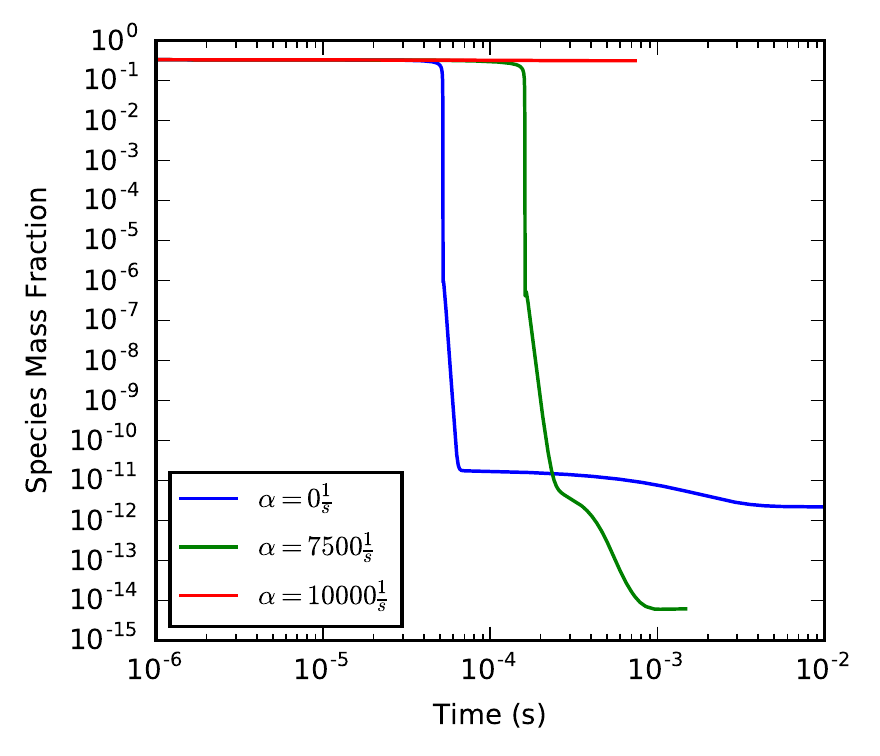} 
		%\caption{}
	\end{subfigure}
	\begin{subfigure}[c]{0.48\textwidth}
	\includegraphics[width=\linewidth]{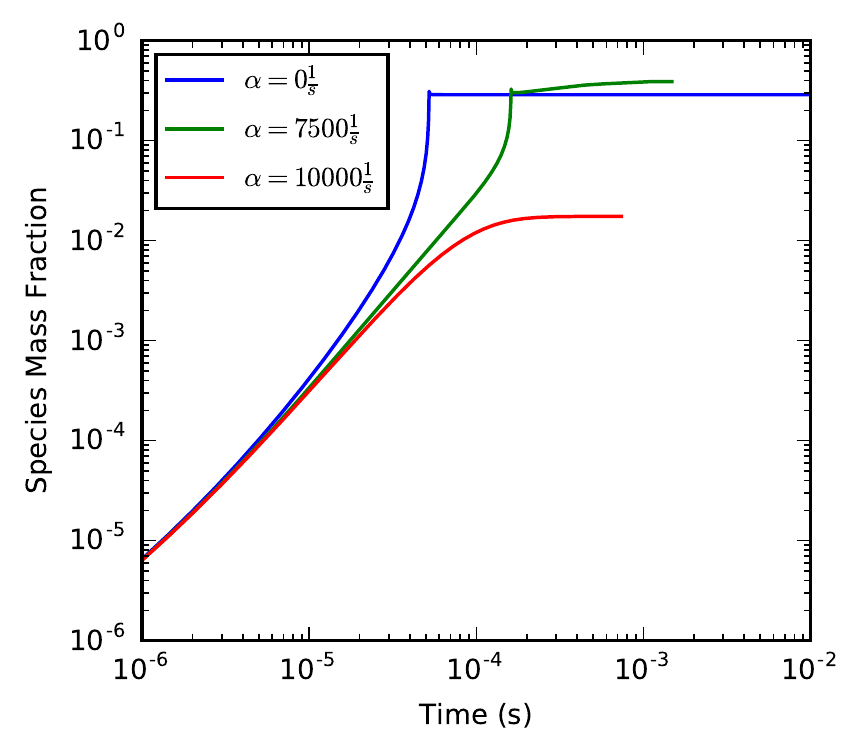} 
	%\caption{}
	\end{subfigure}
	\caption{Results of the integration of a shock wave propagating through a quiescent mixture of stoichiometric methane-oxygen, with an initial pressure and temperature of 4.1 kPa and 293 K, for increasing expansion rates $\alpha = -\frac{1}{v}\frac{Dv}{Dt}$. Mass fraction as a function of time, for a) reactants (CH$_4$), and b) products (H$_2$O).}
	\label{fig:CJPostShockSpecies}
\end{figure}

As an alternative to studying the temperature and thermicity evolutions, one can study the evolution of chemical species in the post-shock region. This can offer an alternative description of ignition and quenching of the post-shock combustion process. Figure \ref{fig:CJPostShockSpecies} a) shows the mass fraction of a reactant (CH$_4$). Studying the solution corresponding to a shock without any expansion, one sees a sharp decline in the mass fraction of the reactants occurring after the induction time. As in Figure \ref{fig:CJPostShock}, the induction time increases with increasing expansion rates between the ignition cases. When increasing the expansion further, the reactants are very minimally depleted, clearly differentiating between the expansion rates associated with ignition and quenching. Similarly, in Figure \ref{fig:CJPostShockSpecies} b), the product mass fraction increases by an order of magnitude when ignition occurs, until plateauing at the final product concentration. The quenched case has an increasing product mass fraction, however the products never constitute the majority of the chemical species in the mixture, plateauing at around 1\% of the overall mass.  

\subsection{1-step Combustion Model}
\label{ch:1StepModel}
The problem of ignition in the presence of volumetric expansion can also be modeled with a single overall reactive step for a perfect gas.  Such a treatment permits to obtain an analytical result for the effect of expansion on ignition delay and the critical expansion rate that can prevent ignition.  The following treatment follows closely the development of \citet{Radulescu&Maxwell2010}, who extended the model of \citet{eckett2000role}. As in the general problem outlined in the previous section, the problem is defined by the energy conservation along a fluid element, the rate of expansion (assumed known \textit{a priori} through the constant $\alpha$) and the evolution of the mass fraction of products $Y$, modeled here with a single step Arrhenius reaction.  In dimensional terms, the initial value problem becomes
\begin{align}
c_v \frac{dT}{dt} &= Q \frac{d Y}{dt} + \frac{p}{\rho^2} \frac{d\rho}{dt},  \nonumber\\ 
\frac{d \ln \rho}{d t}&=-\alpha, \nonumber\\
\rho \frac{d Y}{dt} &= k \rho^n (1-Y)^n \exp \left(-\frac{Ea}{RT}\right),\label{eq:1stpTempCons}
\end{align}
\noindent with initial conditions immediately behind the shock denoted with a subscript $s$:
\begin{align}
T(t=0)=T_s,\; \rho(t=0)=\rho_s \;\textrm{and}\; Y(t=0)=0.
\end{align}
It is worthwhile to non-dimensionalize these equations by choosing the scales $T_s$ as characteristic temperature, $\rho_s$ as characteristic density and the ignition delay obtained in the limit of high activation energy as time scale, i.e.,
\begin{equation}
t_{is}=k^{-1}\rho_s^{-(n-1)} \left(\frac{Ea}{RT_s}\right)^{-1}\left(\frac{Q}{c_vT_s}\right)^{-1}\exp\left(\frac{Ea}{RT_s}\right).
\end{equation}
The latter choice will be justified\textit{ a posteriori} in the following perturbation analysis in the limit of high activation energy.  The new non-dimensional variables and parameters are
\begin{align}
\tilde{T} = \frac{T}{T_s} ,\; \tilde{\rho} = \frac{\rho}{\rho_s},\;  \tilde{t} = \frac{t}{t_{is}},\;  \epsilon=\left(\frac{Ea}{RT_s}\right)^{-1},\; \beta=\frac{Q}{c_vT_s},\; \mathrm{Da}=\frac{1}{\alpha t_{is}},
\end{align}
\noindent where we have introduced the Damkohler number $ \mathrm{Da}$ to denote the ratio of expansion time over the nominal ignition delay time.  This parameter controls the ignition in the resulting non-dimensional initial value problem:  
\begin{align}
\frac{d  \tilde{T} }{d \tilde{t} }&=\beta \frac{d Y}{d \tilde{t} } - \frac{\gamma-1}{\mathrm{Da}} \tilde{T}, \nonumber\\
\frac{d \ln \tilde{\rho}}{d \tilde{t}}&=-\frac{1}{\mathrm{Da}}, \nonumber\\
\frac{d Y}{d\tilde{t}} &= \frac{\epsilon}{\beta} \tilde{\rho}^{n-1} (1-Y)^n \exp \left(\frac{\tilde{T}-1}{\epsilon \tilde{T}}  \right),\nonumber\\ 
\tilde{T}(\tilde{t}=0)=1,\;\tilde{\rho}(\tilde{t}&=0)=1\;\textrm{and}\; Y(\tilde{t}=0)=0. \label{eq:1stepndproblem}
\end{align}
By inspection, the energy equation has the same form as the classical Frank-Kamenetskii homogeneous explosion problem with heat loss \citep{Williams2018}, the loss term being proportional to temperature.  This system is characterized by a critical value of loss rate that can suppress the thermal ignition, parametrized by the Damkohler number.  Although the critical Damkohler number can be determined numerically, as it was performed in the previous section for full chemistry, an analytical approximation can be found in the limit of high activation energy (or small $\epsilon$).  We perturb the initial state by a small perturbation, i.e.,
\begin{align}
\tilde{T} &= 1 + \epsilon\tilde{T}_{II}(\tilde{t})+\mathcal{O}(\epsilon^2), \label{eq:asymexpansionT}\\
Y &= 0 + \epsilon\lambda_{II}(\tilde{t})+\mathcal{O}(\epsilon^2),\\
\tilde{\rho} &= 1 + \epsilon\tilde{\rho}_{II}(\tilde{t})+\mathcal{O}(\epsilon^2). \label{eq:asymexpansionrho}
\end{align}
We substitute these expansions in the original problem \eqref{eq:1stepndproblem} and solve at the various orders.  The $\mathcal{O}(1)$ problem is the constant state at the initial condition.  The $\mathcal{O}(\epsilon)$ problem becomes:
\begin{align}
\frac{d  \tilde{T_{II}} }{d \tilde{t} }= \exp\left( \tilde{T}_{II}\right) - \zeta,\; \tilde{T}_{II}(\tilde{t}=0)=0, \label{eq:1stepndproblemOeps}
\end{align}
\noindent with 
\begin{equation}
\zeta\equiv\frac{1}{\epsilon}\frac{\gamma-1}{\mathrm{Da}} =\mathcal{O}(1).\label{eq:zetadef}
\end{equation} 
\noindent This can be integrated to yield the solution to the temperature perturbation in the induction zone as 
\begin{equation}
\tilde{T}_{II}(\tilde{t}) = \ln \left(\frac{\zeta}{1-(1-\zeta)e^{\tilde{t}\zeta}}\right).\label{eq:tperturb1step}
\end{equation}
Thermal ignition is associated with the manifestation of $\tilde{T}_{II} \rightarrow \infty$ as $\tilde{t} \rightarrow \tilde{t}_i$.  Taking this limit in  \eqref{eq:tperturb1step}, we obtain an expression for the ignition delay $\tilde{t}_i$ in the presence of expansion
\begin{equation}
\tilde{t}_i=\ln \left(\left(\frac{1}{1-\zeta}\right)^{\frac{1}{\zeta}} \right). \label{eq:ignitiondelay1step}
\end{equation}
Ignition occurs in finite time only if $\zeta < 1$.  The critical value of $\zeta$, henceforth denoted with a $\ast$ as $\zeta^\ast$, is 1.  This is the condition for ignition in the presence of expansion.  Note that ignition occurs at $\tilde{t}_i=1$ when there is no expansion ($\zeta=0$), hence justifying \textit{a posteriori} our choice for time scale.  

In dimensional variables, the condition for thermal ignition $\zeta < \zeta^\ast=1$ is 
\begin{equation}
\zeta = \frac{1}{\epsilon}\frac{\gamma-1}{\mathrm{Da}} =\frac{Ea}{RT_s}\left(\gamma-1 \right)\alpha t_{is} < 1. \label{eq:igcriterion1step}
\end{equation}

The comparison between the prediction of this analysis with the numerical integration of the full multi-specie problem will be discussed later in this chapter. 

\subsection{2-Step Combustion Model}
\label{ch:2stepModel}
As discussed in the introduction, critical ignition has been also addressed by Lundstrom and Oppenheim using a generalized ignition delay model.  In the modern literature, the framework of their analysis falls within the scope of a two-step chemical model that is characterized by a thermally neutral induction zone followed by an exothermic stage.  This structure does not necessarily associate the ignition with thermal explosion, but relies on either simulations or experiments to determine the effective kinetics of the induction delay time. The kinetics of the induction zone can also model a radical explosion in chain-branching systems.  In any case, it is worthwhile formulating a two-step model for ignition with expansion and compare its predictions for the ignition delay variation with expansion rate and the critical conditions for ignition with the full chemistry and one-step model discussed above. 

In the two-step model, the sequential induction and reaction zones are modelled by the rate equations
\begin{align}
\frac{D\xi}{Dt} &= k_i \rho^{n_i-1} \exp \left(-\frac{E_a}{RT} \right), \label{eq:2stepInduction}\\
\frac{D Y}{Dt} &= k_r \rho^{n_r-1} (1-Y)^{n_r} H(\xi), \label{eq:1stpRate}
\end{align}
\noindent where $\xi$ is the progress variable of the induction zone, ranging from 0 to 1 at the end of the induction time and $Y$ is the mass fraction of product, ranging from 0 to 1 in the main reaction zone.  The reaction zone only starts when the progress variable $\xi$ of the induction zone reaches unity; $H(\xi)$ is the switch:
\begin{align}
H(\xi) &= 
\begin{cases}
0 &\text{for } \xi < 1\\
1 &\text{for } \xi \geq 1\\
\end{cases}
\end{align} 
The parameters of the two step model are the rate constants rate multipliers $k_i$ and $k_r$, the reaction orders $n_i$ and $n_r$ and the activation temperature of the induction zone $Ea/R$.   

The same energy equation \eqref{eq:1stpTempCons} and expansion rate hold as for the 1-step model discussed previously: 
\begin{align}
c_v \frac{dT}{dt} &= Q \frac{d Y}{dt} + \frac{p}{\rho^2} \frac{d\rho}{dt}, \\
\frac{d \ln \rho}{d t}&=-\alpha, 
\end{align}
\noindent with initial conditions immediately behind the shock denoted with a subscript $s$:
\begin{align}
T(t=0)=T_s,\; \rho(t=0)=\rho_s, \; \xi(t=0)=0,\;Y(t=0)=0.
\end{align}
We again non-dimensionalize these equations by choosing the scales $T_s$ as characteristic temperature, $\rho_s$ as characteristic density and the ignition delay without expansion, which for the two-step model, 
\begin{equation}
t_{is}=k_i^{-1}\rho_s^{-(n_i-1)} \exp\left(\frac{Ea}{RT_s}\right).
\end{equation}
\noindent The new non-dimensional variables and parameters are
\begin{align}
\tilde{T} = \frac{T}{T_s} ,\; \tilde{\rho} = \frac{\rho}{\rho_s},\;  \tilde{t} = \frac{t}{t_{is}},\;  \epsilon=\left(\frac{Ea}{RT_s}\right)^{-1}, \nonumber\\
 \beta=\frac{Q}{c_vT_s},\; \mathrm{Da}=\frac{1}{\alpha t_{is}},\; \zeta=\frac{\gamma-1}{\epsilon \mathrm{Da}},\; \tilde{k}_r=k_rt_{is}\rho_s^{n_r-1}.
\end{align}
\noindent The resulting non-dimensional initial value problem becomes:  
\begin{align}
\frac{d  \tilde{T} }{d \tilde{t} }&=\beta \frac{d Y}{d \tilde{t} } - \epsilon \zeta \tilde{T}, \nonumber\\
\frac{d \ln \tilde{\rho}}{d \tilde{t}}&=-\epsilon\frac{\zeta}{\gamma-1}, \nonumber\\
\frac{d\xi}{dt}&= \tilde{\rho}^{n_i-1} \exp \left(\frac{\tilde{T}-1}{\epsilon \tilde{T}}  \right),\nonumber\\ 
\frac{d Y}{d\tilde{t}} &= \tilde{k}_r \tilde{\rho}^{n_r-1} (1-Y)^{n_r}H(\xi), \nonumber\\ 
\tilde{T}(\tilde{t}=0)=1,\;\tilde{\rho}(\tilde{t}&=0)=1\; \xi(t=0)=0\;,Y(\tilde{t}=0)=0. \label{eq:2stepndproblem}
\end{align}
In the induction zone, there is no heat release and the problem is further simplified:
\begin{align}
\frac{d  \tilde{T} }{d \tilde{t} }&=- \epsilon \zeta \tilde{T}, \nonumber\\
\frac{d \ln \tilde{\rho}}{d \tilde{t}}&=-\epsilon\frac{\zeta}{\gamma-1}, \nonumber\\
\frac{d\xi}{dt}&= \tilde{\rho}^{n_i-1} \exp \left(\frac{\tilde{T}-1}{\epsilon \tilde{T}}  \right),\nonumber\\ 
\tilde{T}(\tilde{t}=0)=1,\;\tilde{\rho}(\tilde{t}&=0)=1,\; \xi(t=0)=0. \label{eq:2stepndprobleminduction}
\end{align}
These ODEs are readily integrated from the shock to the end of the induction zone, marking the ignition time $\tilde{t}=\tilde{t}_i$:
\begin{align}
\int_0^1 d\xi=1&=\int_0^{\tilde{t}_i}\tilde{\rho}^{n_i-1} \exp \left(\frac{\tilde{T}-1}{\epsilon \tilde{T}}\right) d\tilde{t},  \label{eq:ti2step}
\end{align}
\noindent with $\tilde{\rho}(\tilde{t})$ and $\tilde{T}(\tilde{t})$ in the last integral given by 
\begin{align}
\tilde{\rho}&=\exp\left( -\epsilon\frac{\zeta}{\gamma-1}\tilde{t}\right), \\
\tilde{T}&=\exp\left( -\epsilon\zeta\tilde{t}\right).
\end{align}
This expression is an implicit relation that defines the ignition delay time in the presence of expansion, which can be evaluated numerically. 
\subsubsection*{Reaction order $n_i =1$}
Analytical expressions can be obtained for a reaction order of $n_i =1$.  The implicit expression \eqref{eq:ti2step} for $\tilde{t_i}$ becomes:
\begin{equation}
\zeta=\frac{1}{\epsilon}\exp\left(\frac{1}{\epsilon}\right) \left(  -\mathrm{Ei} \left(-\frac{1}{\epsilon}\right)+\mathrm{Ei} \left(-\exp\left(  \epsilon \zeta \tilde{t}_i \right)     \right) \right),   \label{eq:implicittin1}
\end{equation}
\noindent where $\mathrm{Ei}$ is the exponential integral function. 
The ignition limit corresponds to letting $\tilde{t}_i\rightarrow\infty$, yielding the critical condition for ignition 
\begin{equation}
\zeta^\ast=\frac{\gamma-1}{\epsilon\mathrm{Da}^\ast}=-\frac{1}{\epsilon}\exp\left(\frac{1}{\epsilon}\right) \mathrm{Ei} \left(-\frac{1}{\epsilon}\right).
\end{equation}
The asymptotic expansion of this expression can also be written as
\begin{equation}
\zeta^\ast=\frac{\gamma-1}{\epsilon\mathrm{Da}^\ast}=1-\epsilon+\mathcal{O}\left(\epsilon^2 \right).
\end{equation}
To order $\mathcal{O}(1)$, the limit agrees with the result obtained for the 1 step model given by \eqref{eq:igcriterion1step}.

\subsubsection{Ignition with the two step model with $\zeta=\mathcal{O}(1)$}
The analysis showed that the two step model limit expressed in terms of $\zeta$ was in agreement with that of the 1-step model to $\mathcal{O}(1)$, yielding a critical ignition limit given by $\zeta^\ast=1$ .  It is worthwhile to further pursue this limit and determine the ignition delay predicted by the two-step model in the presence of expansion.  For the one step model, the analysis required that $\zeta=\mathcal{O}(1)$ for balancing terms in the $\mathcal{O}(\epsilon)$ problem.  It is straightforward to show that taking the expansions \eqref{eq:asymexpansionT} and \eqref{eq:asymexpansionrho} to solve the problem given by \eqref{eq:2stepndprobleminduction}, the leading order solution for the ignition delay requires $\zeta=\mathcal{O}(1)$ and $\gamma-1 =\mathcal{O}(1)$ for matching and yields the same result for the ignition delay as the 1-step model, namely \eqref{eq:ignitiondelay1step}. 

At $\mathcal{O}(\epsilon)$, we obtain: 
\begin{align}
\tilde{T}_{II} &= -\zeta\tilde{t}, \\
\tilde{\rho}_{II} &= -\frac{\zeta}{\gamma-1}\tilde{t}.
\end{align}
\noindent resulting in the asymptotic approximations for temperature and density in the induction zone given by
\begin{align}
\tilde{T} &=1 - \epsilon \zeta\tilde{t} +\mathcal{O}(\epsilon^2), \\
\tilde{\rho} &= 1 - \epsilon \frac{\zeta}{\gamma-1}\tilde{t}+\mathcal{O}(\epsilon^2).
\end{align}
The ignition delay is obtained by substituting these expressions in the integral expression given by \eqref{eq:ti2step}, yielding:
\begin{equation}
1=\int_0^{\tilde{t}_i}  \left(1 - \epsilon (n_i-1)\frac{\zeta}{\gamma-1}\tilde{t}+... \right) \exp \left(- \zeta \tilde{t} \right) d\tilde{t}
\end{equation}
Evaluating the integral for $n_i = 1$ yields the same result as for the 1 step model given by \eqref{eq:ignitiondelay1step}, namely
\begin{equation}
\tilde{t}_i=\ln \left(\left(\frac{1}{1-\zeta}\right)^{\frac{1}{\zeta}} \right). \label{eq:ignitiondelay2stepapprox}
\end{equation}

For $n_i \neq 1$, $\tilde{t}_i$ cannot be obtained in closed form.  However, the ignition limit can be evaluated by taking the limit $\tilde{t}_i \rightarrow \infty$.  Evaluating the integral, the condition for ignition becomes:
\begin{equation}
\zeta^\ast=\frac{\gamma-1}{\epsilon \mathrm{Da}^\ast}=1-\epsilon\frac{n_i-1}{\gamma-1}.
\end{equation}
The non-unity reaction order appears as a perturbation to the $\zeta^\ast=1$ limit obtained above for the 1 step and 2 step models at leading order. 

\subsection{Comparison of the Simplified Models}
\label{ch:ModelComparison}
\begin{figure}
	\begin{subfigure}[c]{0.48\textwidth}
		\includegraphics[width=\linewidth]{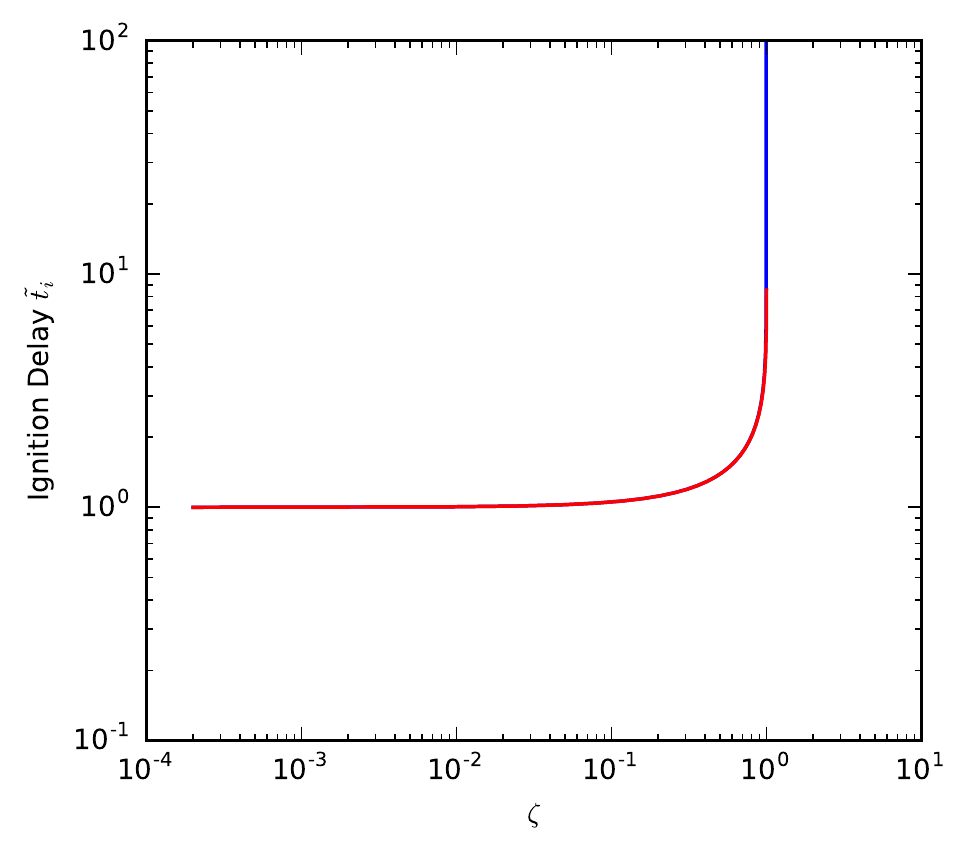} 
		\caption{}
	\end{subfigure}
	\begin{subfigure}[c]{0.48\textwidth}
		\includegraphics[width=\linewidth]{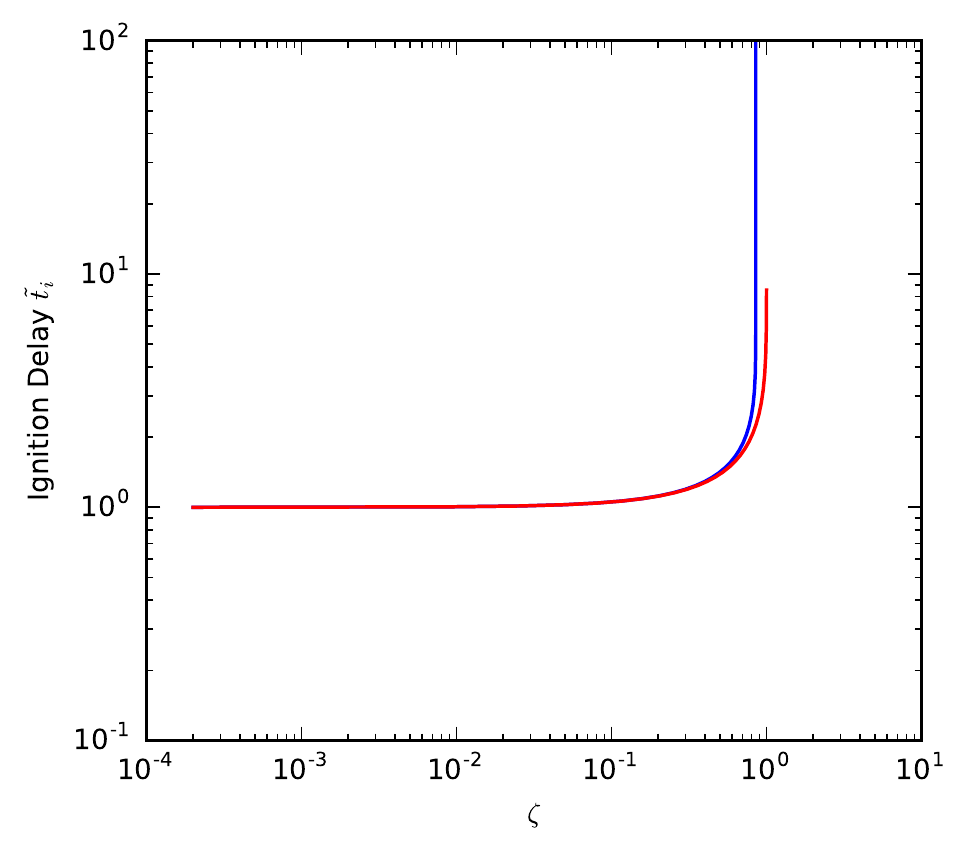} 
		\caption{}
	\end{subfigure}
	\caption{Plot of the non-dimensional ignition $\tilde{t}_{ig}$ as a function of $\zeta$, for the cases of a) $\epsilon = 0.005$ and b) $\epsilon = 0.2$. One notices deviation between the ignition delay predicted by simplified models when $\epsilon$ becomes large. Blue : Implicit 2-step solution given in (\ref{eq:implicittin1}) for $n_i = \mathcal{O}(1)$. Red : 1-step and 2-step solutions in the limit of $\zeta = \mathcal{O}(1)$ given in (\ref{eq:ignitiondelay1step}) and (\ref{eq:ignitiondelay2stepapprox}).  }
	\label{fig:NonDimIgnition}
\end{figure}

The models derived previously in this chapter predict quenching in the limit of $\zeta \simeq 1$. Figure \ref{fig:NonDimIgnition} plots the ignition delay as a function of the controlling parameter $\zeta$, while considering the effect of the magnitude of the inverse activation energy $\epsilon$. This allows us to study the effect of expansion in the induction zone on the predicted ignition delay obtained using simplified models.

Two different solutions modelling the evolution of the ignition delay are compared in Figure  \ref{fig:NonDimIgnition}. The first solution is derived from the two-step model under the assumption that the reaction order of the induction stage is $n_i = 1$, giving the ignition delay implicitly by  (\ref{eq:implicittin1}) for $n_i = 1$. The third solution is obtained independently using the 1-step and 2-step combustion model. After a derivation following the development of Radulescu \& Maxwell, who in turn extended the model of Eckett and co-workers, a solution is obtained under the assumption of a 1-step combustion model in the limit of high activation energy, shown in (\ref{eq:ignitiondelay1step}). An identical expression for the ignition delay is obtained from an asymptotic approximation of the temperature and density in the induction zone of a 2-step chemistry model under the assumption that $\zeta = \mathcal{O}(1)$ and $\gamma-1 = \mathcal{O}(1)$, shown in (\ref{eq:ignitiondelay2stepapprox}).

In the limit of a small inverse activation energy as shown in Figure \ref{fig:NonDimIgnition} a), one notices the very good agreement between the predicted evolution of the ignition delay calculated using the previously presented solutions, including the critical ignition limit when $\zeta = 1$. Conversely, the predictions diverge when $\epsilon$ becomes large, as seen in Figure \ref{fig:NonDimIgnition} b). Studying the evolution of the implicit 2-step solution given in (\ref{eq:implicittin1}), one notices that this solution correctly recovers the expected behaviour when the volumetric expansion rate is small, be it $\tilde{t}_i = t_{ig}/t_{is} \approx 1$. When increasing the controlling parameter, and thus the post-shock volumetric expansion rate, the ignition delay also increases and reaches the critical ignition limit sooner than the other simplified ignition models, at $\zeta\approx0.85$. The solution obtained for 1-step and 2-step chemistry models under the assumption that $\zeta = \mathcal{O}(1)$ reproduces the expected behaviour for small values of the controlling parameter, however also recovers the critical ignition limit specified by $\zeta = 1$.

\subsection{Comparison of Simplified Models with Detailed Chemistry}
\label{ch:DetailedComparison}
Having shown general agreement between the ignition delay predicted by the simplified models derived above, it is now of interest to compare these models with an integration of a post-shock state using detailed chemistry. To do this, we will study ignition behind a decaying shock wave propagating at its CJ velocity following the methodology proposed in sections \ref{ch:IgnitionDescription} and \ref{ch:CanteraDescription}. The lead shock will be subjected to increasing post-shock volumetric expansion rates ($\alpha$) until the quenching of post-shock reactions occurs due to expansion cooling. The results obtained from the detailed chemical model can be compared with the simplified models by calculating the controlling parameter, 
\begin{equation}
\zeta = \left(\gamma-1\right) \frac{E_a}{RT_s}\alpha t_{is},
\end{equation}
for each expansion rate studied. The expansion rate $\alpha$ is the only parameter in this equation which changes, as $\gamma$, $E_a$, $T_s$, and $t_{is}$ are all properties of the shock and will thus only vary with the shock speed. The activation energy is calculated as the slope of an Arrhenius plot of the ignition delay against the inverse of the temperature, while keeping density constant, 
\begin{equation}
Ea = \frac{\log{t_{ig}^+}-\log{t_{ig}^-}}{\log{T_s^+}-\log{T_s^-}},
\end{equation}
with $T_s^\pm$ a perturbation of the post-shock temperature at constant density, and $t_{ig}^\pm$ the ignition delay associated with its respective temperature perturbation. 

\begin{figure}
	\begin{subfigure}[c]{0.48\textwidth}
		\includegraphics[width=\linewidth]{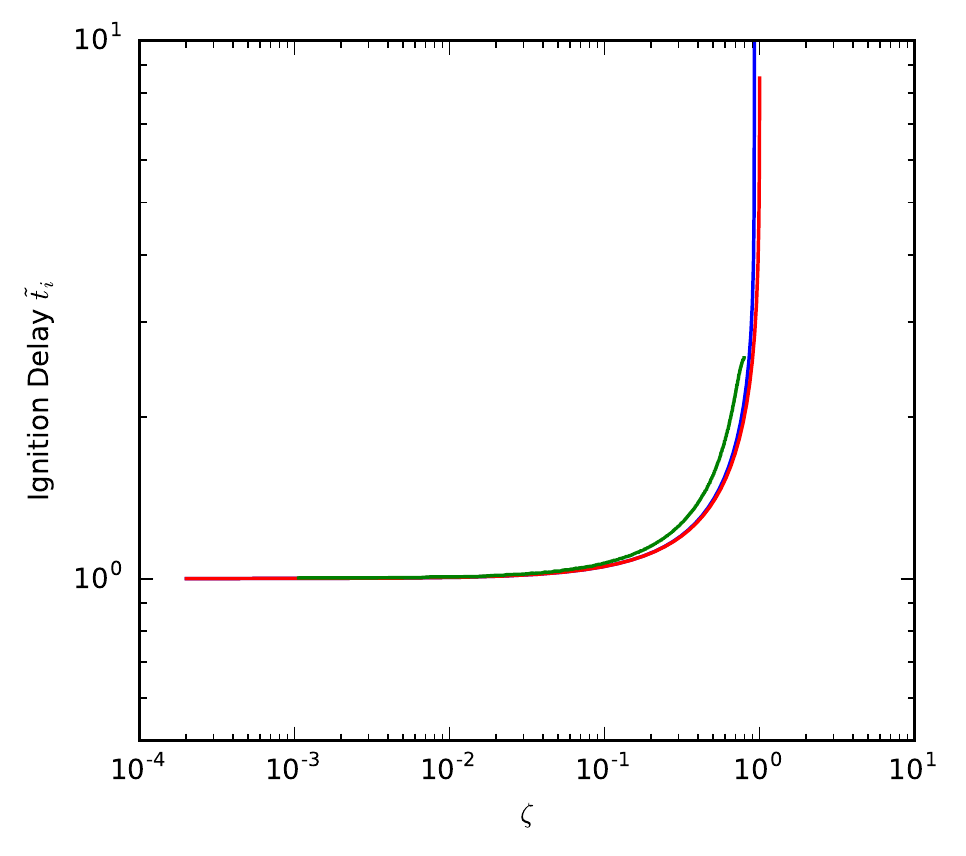} 
		\caption{}
	\end{subfigure}
	\begin{subfigure}[c]{0.48\textwidth}
		\includegraphics[width=\linewidth]{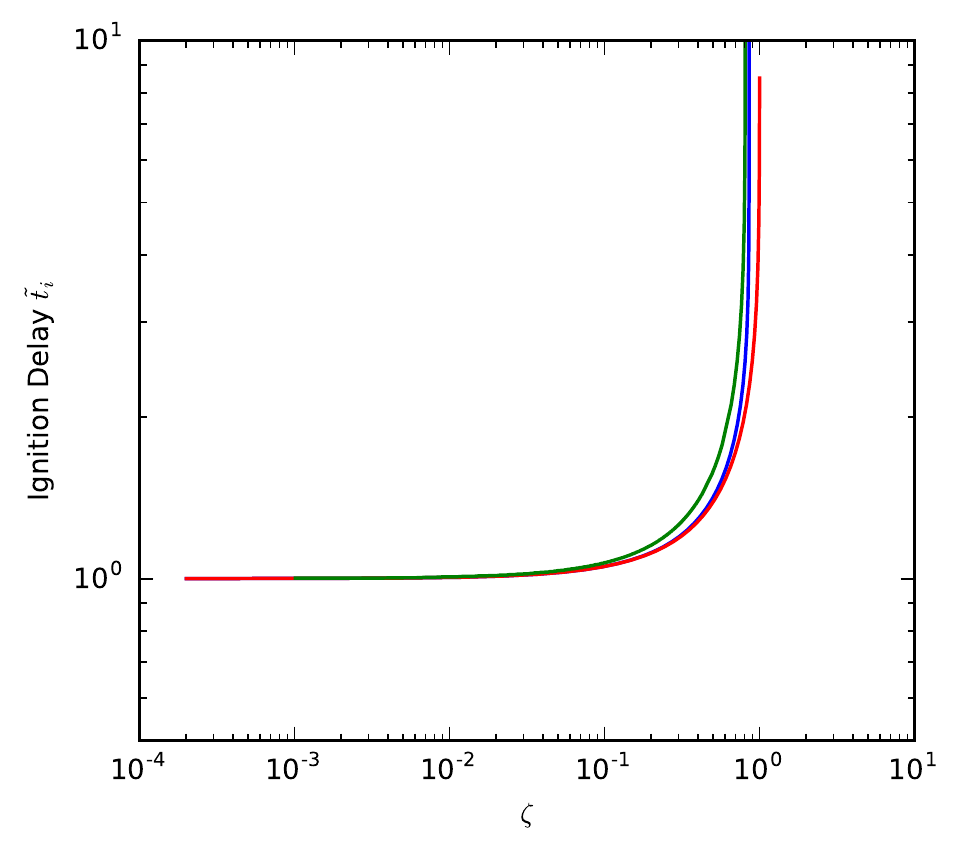} 
		\caption{}
	\end{subfigure}
	\caption{The evolution of the non-dimensional ignition delay $\tilde{t}_{ig}$ as a function of $\zeta$ behind a shock wave propagating at the CJ velocity through a quiescent mixture of a) 2H$_2$ + O$_2$ + 7Ar with $p_0 = 4.1$ kPa ($\epsilon = 0.089$), and b) CH$_4$ + 2O$_2$ with $p_0 = 3.5$ kPa ($\epsilon = 0.194$). One notices the simplified models recover the evolution of the ignition delay obtained using full chemistry. Blue : Implicit 2-step solution given in (\ref{eq:implicittin1}) for $n_i = \mathcal{O}(1)$. Red : 1-step and 2-step solutions in the limit of $\zeta = \mathcal{O}(1)$ given in (\ref{eq:ignitiondelay1step}) and (\ref{eq:ignitiondelay2stepapprox}). Green : solution obtained by integrating the post-shock state using a detailed chemical model.}
	\label{fig:NonDimIgnition_DetailedChemistry}
\end{figure}

The ignition delay $t_{ig}$ normalized by the ignition delay without expansion, $t_{is}$, is shown in Figure \ref{fig:NonDimIgnition} as a function of the controlling parameter. In the limit of small expansion rates, the predicted ignition delay found using simplified models agree with those found using detailed chemistry. There is excellent agreement when studying the critical ignition which occurs as $\zeta\rightarrow1$. As seen in Figure \ref{fig:NonDimIgnition} b), the implicit 2-step model has a critical ignition asymptote associated with a smaller value of $\zeta$ than the other models. This behaviour is also seen in critical ignition calculated using detailed chemistry, occurring at a smaller value of $\zeta\approx0.8$ in Figure \ref{fig:NonDimIgnition_DetailedChemistry}. The implicit 2-step model therefore recovers the ignition time predicted by detailed chemistry the best of any simplified model derived earlier in this chapter. Nevertheless, the models derived in the limit of $\zeta = \mathcal{O}(1)$ for 1-step and 2-step chemistry also recover critical ignition as $\zeta\rightarrow1$. As seen in Figure \ref{fig:NonDimIgnition_DetailedChemistry}, this criterion agrees with the critical ignition calculated using detailed chemistry.

\subsection{Criterion for Ignition Behind Decaying Shock Waves}
\label{ch:IgnitionCriterion}
To summarize, we have now formulated a criterion for the critical ignition, which yields approximately the same result for both 1 step thermal explosions, 2 step models for chain-branching explosions.   This criterion, derived in sections \ref{ch:1StepModel} and \ref{ch:2stepModel}  
\begin{equation}
\zeta = \frac{1}{\epsilon}\frac{\gamma-1}{\mathrm{Da}} < 1, 
\end{equation}
was also found in very good agreement with the results of the integration using a realistic network of chemical reactions. This criterion can be recast in dimensional form in terms of thermo-chemical properties of the gas at the post-shock state and the post-shock volumetric expansion rate,
\begin{equation}
\frac{Ea}{RT_s}\left(\gamma-1 \right)\alpha t_{is}(T_s) < 1. 
\end{equation}
%This form can be used to accurately predict the quenching occurring in Figure \ref{fig:NonDimIgnition_DetailedChemistry}, with $E_a$, $T_s$, and $\gamma$ a function of the speed of the lead shock. 
Making use of the relation between expansion rate along the particle and the shock dynamics, obtained via the shock change equations \eqref{eq:StrongShockChange}, this criterion can be re-written for a strong shock in a perfect gas as: 
\begin{equation}
\frac{E_a}{RT_s}\left(\gamma-1\right) t_{is}(T_s) \left(\frac{-6}{\gamma+ 1}\frac{\dot{D}}{D} - 2D\kappa\frac{\gamma-1}{\left(\gamma+1\right)^2}\right) < 1.  \label{eq:TheCriterion}
\end{equation}
This criterion links the quenching of post-shock reactions solely to the dynamics of the lead shock. From the shock speed and curvature, every term in this equation can be calculated knowing the quiescent mixture through which the shock propagates. As shown in section \ref{ch:DetailedComparison}, this criterion can be used to model the quenching of reactions behind a shock due to expansion cooling and recovers the ignition limit calculated using detailed chemistry models.

The criterion given by \eqref{eq:TheCriterion} is identical with the one derived by Eckett and co-workers by a very different approach for a 1 step model, and by Radulescu \& Maxwell; our result also includes the effect of curvature neglected by these previous authors.   The derivation of the models presented in sections \ref{ch:1StepModel} and \ref{ch:2stepModel} unified Eckett's 1-step Critical Decay Rate model and Lundstrom \& Oppenheim's treatment of induction kinetics and Taylor Sedov blast waves.  In the limit of $\gamma - 1 = \mathcal{O}(1)$ and high activation energy, an identical solution is obtained from these two independent treatments of post-shock ignition. It was also found that the models yielding an explicit expression for the ignition delay also recovered the behaviour of the implicit 2-step solution in the limit of small $\epsilon$, with the predictions diverging as $\epsilon$ ceases to be small. The ignition delay predicted using models derived from simplified chemical kinetics was finally compared to that calculated using detailed chemistry, and general agreement was observed. The critical ignition asymptote calculated using detailed chemistry agreed particularly well with the one predicted by the implicit 2-step model yet remains compatible with the ignition delay predicted using the models who give an explicit expression for the ignition delay, notably the approximate 2-step linearization of the implicit model in $\epsilon$ and the model obtained in the limit of $\zeta = \mathcal{O}(1)$. A strong case can thus be made for the usage of the ignition criterion presented in section (\ref{eq:TheCriterion}), as it is found to offer a good prediction of quenching caused by expansion cooling. 
\section{Experimental Setup and Apparatus}

\begin{figure}
	\centering
	\includegraphics[width=\textwidth]{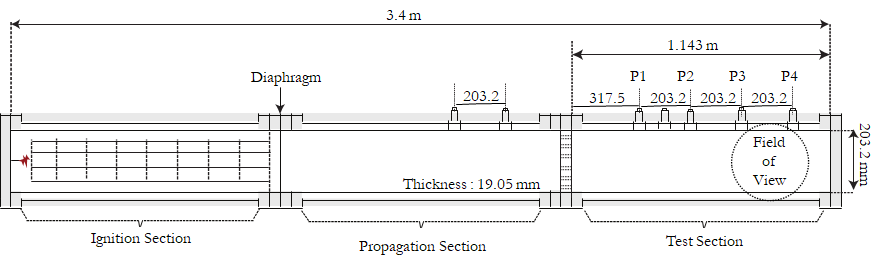}
	\caption{Schematic drawing of the shock tube.}
	\label{fig:ShockTube}
\end{figure}

The experiments were performed in a thin shock tube with a rectangular cross-section, as sketched in Figure \ref{fig:ShockTube}. The shock tube is 3.4 m long with a height of 203 mm and a depth of 19 mm, allowing for the visualization of the flow field while minimizing gasdynamic effects in the 3rd dimension. The tube is composed of an ignition, a propagation, and a test section of equal lengths, with aluminium panels placed on the ignition and propagation sections and non-tempered sodium silicate panels placed on the test section to allow visualization of the flow. More details pertaining to the shock tube can be found in \citet{ShockTube}. 

A high voltage igniter was used to initiate the gas by the means of a spark plug installed in the end-wall of the ignition section. The two 1$\mu$F capacitors within this system can be charged to a voltage of 30 kV, and subsequently discharged to deposit a charge of approximately 1 kJ with a deposition time of under 2 $\mu$s. A mesh grid placed within the ignition section further promotes the onset of detonation.

The experiments are visualized using Z-type Schlieren photography as highlighted in \citet{Settles2013Book}. This setup is achieved using two parabolic mirrors with a 2540 mm focal length, a 360 W light source, and a high-speed camera with a frame rate of 77481 fps and a resolution of 384 pixels by 288 pixels. The interframe time is 12.91 $\mu$s and the exposure time is 0.468 $\mu$s.

The initial conditions of the two premixed combustible mixtures are shown in Table \ref{tab:InitConditions}. These conditions are chosen such that only a single detonation cell forms across the height of the shock tube. The test conditions are the same as those used by \citet{Maxwell2017} for stoichiometric methane-oxygen experiments, and \citet{Xiao2020} for argon-diluted hydrogen-oxygen experiments. A driver gas mixture of C$_2$H$_4$ + 3O$_2$ was used to help facilitate the onset of detonation, with initial conditions of $p_0 = 10.34$ kPa ($1.5$ psi), and $T_0 = 293$ K. Disposable aluminium diaphragms were placed between the initiation and propagation sections to partition the gases as shown in Figure \ref{fig:ShockTube}. 

\begin{table}
	\centering
	\caption{Shock tube experimental test conditions.}
	\vspace{-0.15cm}
	\label{tab:InitConditions}
	\begin{tabular}{c c c c c c}
		\hline \hline
		Mixture  &$p_0$ (kPa) & $T_0$ (K) & Interframe Time ($\mu$s) & $\bar{D}$ (m/s) & $D_{CJ}$ (m/s)\\
		\hline
		CH$_4$ + 2O$_2$ & $3.5$ & $293$ & $12.97$ & $1870$ & $2244$ \\
		2H$_2$ + O$_2$ + 7Ar & $4.1$ & $293$ & $12.91$ & $1340$ & $1603$ \\ 
		\hline
	\end{tabular}
\end{table}

The detonable gaseous mixtures were prepared using the method of partial pressures. The mixing tank was first evacuated to a pressure of 40 Pa, then filled using the method of partial pressures, and remained in the mixing day for at least a day before use. Before each experiment, the shock tube is emptied by a vacuum pump to a pressure of at most 80 Pa before filling the driver section followed by the driven section.

\section{High Speed Visualization Results}
\label{ch:Visualization}
\begin{figure}
	\centering
	\begin{subfigure}[c]{0.32\textwidth}
		\includegraphics[width=\linewidth]{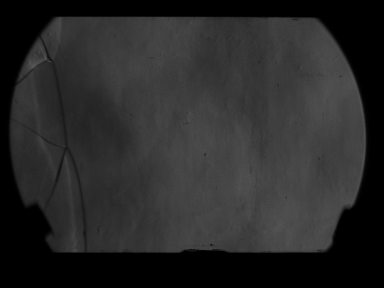} 
		%\caption{}
	\end{subfigure}
	\begin{subfigure}[c]{0.32\textwidth}
		\includegraphics[width=\linewidth]{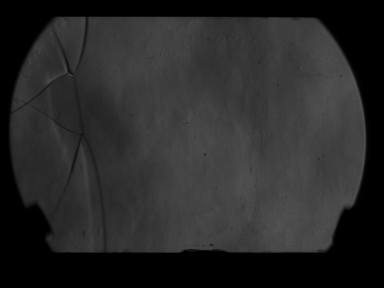} 
		%\caption{}
	\end{subfigure}
	\begin{subfigure}[c]{0.32\textwidth}
		\includegraphics[width=\linewidth]{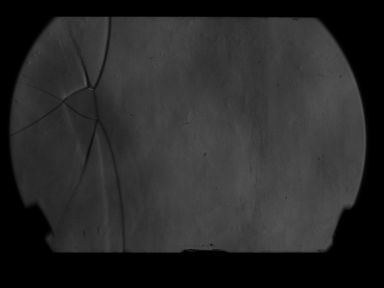} 
		%\caption{}
	\end{subfigure}

	\begin{subfigure}[c]{0.32\textwidth}
		\includegraphics[width=\linewidth]{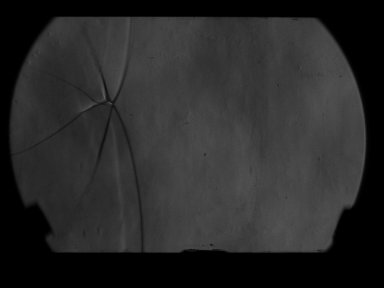} 
		%\caption{}
	\end{subfigure}
	\begin{subfigure}[c]{0.32\textwidth}
		\includegraphics[width=\linewidth]{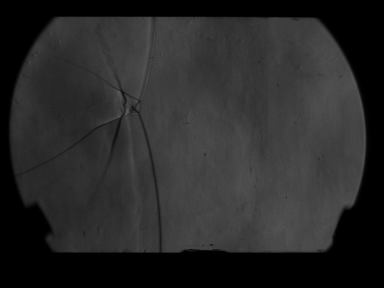} 
		%\caption{}
	\end{subfigure}
	\begin{subfigure}[c]{0.32\textwidth}
		\includegraphics[width=\linewidth]{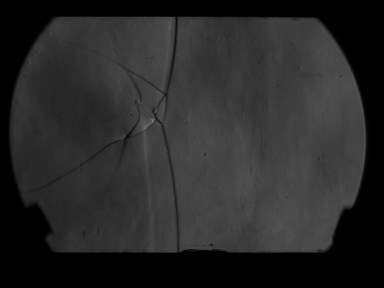} 
		%\caption{}
	\end{subfigure}

	\begin{subfigure}[c]{0.32\textwidth}
		\includegraphics[width=\linewidth]{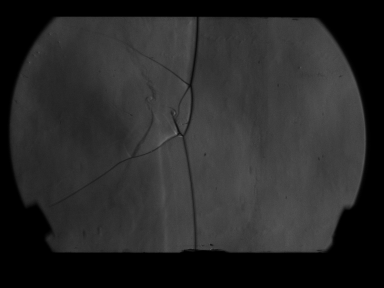} 
		%\caption{}
	\end{subfigure}
	\begin{subfigure}[c]{0.32\textwidth}
		\includegraphics[width=\linewidth]{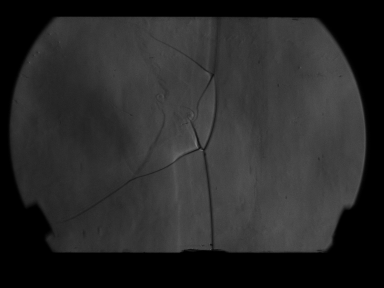} 
		%\caption{}
	\end{subfigure}
	\begin{subfigure}[c]{0.32\textwidth}
		\includegraphics[width=\linewidth]{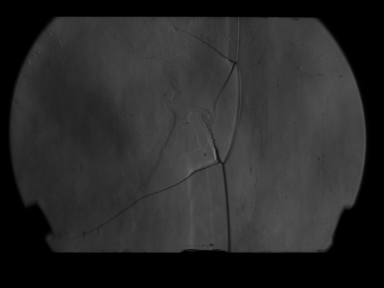} 
		%\caption{}
	\end{subfigure}

	\begin{subfigure}[c]{0.32\textwidth}
		\includegraphics[width=\linewidth]{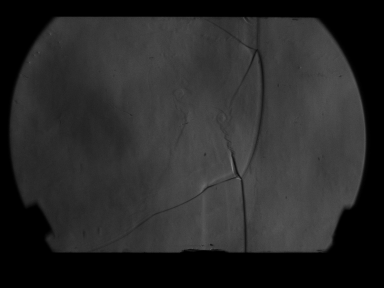} 
		%\caption{}
	\end{subfigure}
	\begin{subfigure}[c]{0.32\textwidth}
		\includegraphics[width=\linewidth]{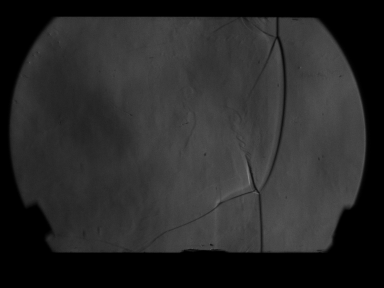} 
		%\caption{}
	\end{subfigure}
	\begin{subfigure}[c]{0.32\textwidth}
		\includegraphics[width=\linewidth]{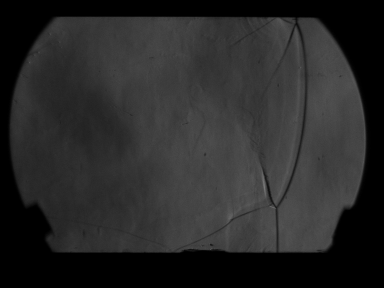} 
		%\caption{}
	\end{subfigure}

	\begin{subfigure}[c]{0.32\textwidth}
		\includegraphics[width=\linewidth]{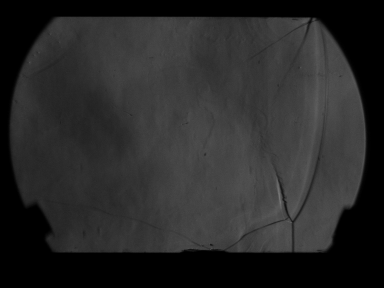} 
		%\caption{}
	\end{subfigure}
	\begin{subfigure}[c]{0.32\textwidth}
		\includegraphics[width=\linewidth]{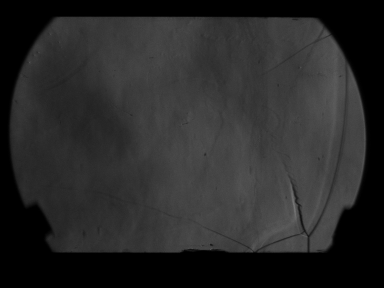} 
		%\caption{}
	\end{subfigure}
	\begin{subfigure}[c]{0.32\textwidth}
		\includegraphics[width=\linewidth]{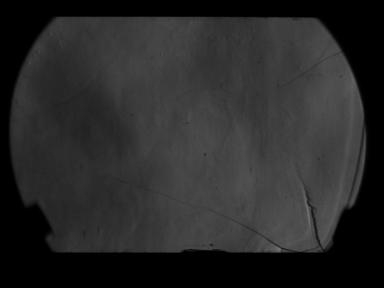} 
		%\caption{}
	\end{subfigure}
	\caption{Schlieren visualization of the experiment 13\_H2 conducted in 2H$_2$ + O$_2$ + 7Ar. }
	\label{fig:HydrogenExpSummary}
\end{figure}

\begin{figure}
	\centering
	\begin{subfigure}[c]{0.32\textwidth}
		\includegraphics[width=\linewidth]{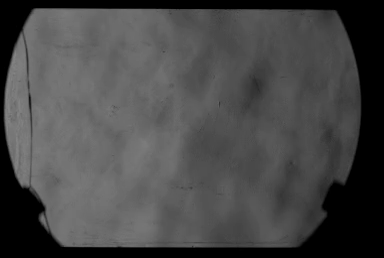} 
		%\caption{}
	\end{subfigure}
	\begin{subfigure}[c]{0.32\textwidth}
		\includegraphics[width=\linewidth]{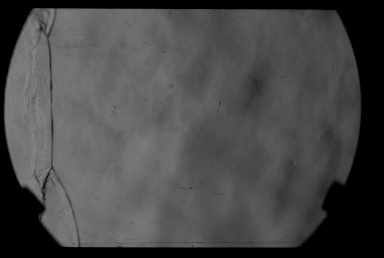} 
		%\caption{}
	\end{subfigure}
	\begin{subfigure}[c]{0.32\textwidth}
		\includegraphics[width=\linewidth]{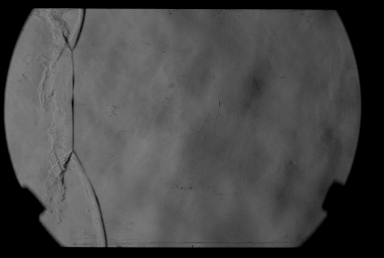} 
		%\caption{}
	\end{subfigure}
	
	\begin{subfigure}[c]{0.32\textwidth}
		\includegraphics[width=\linewidth]{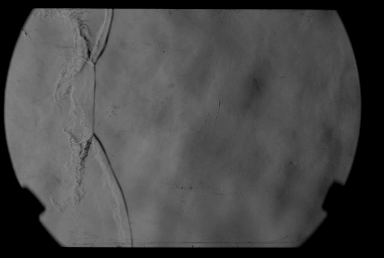} 
		%\caption{}
	\end{subfigure}
	\begin{subfigure}[c]{0.32\textwidth}
		\includegraphics[width=\linewidth]{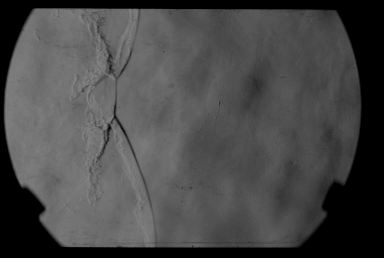} 
		%\caption{}
	\end{subfigure}
	\begin{subfigure}[c]{0.32\textwidth}
		\includegraphics[width=\linewidth]{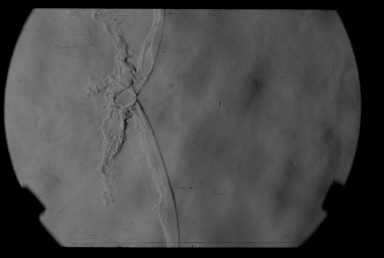} 
		%\caption{}
	\end{subfigure}
	
	\begin{subfigure}[c]{0.32\textwidth}
		\includegraphics[width=\linewidth]{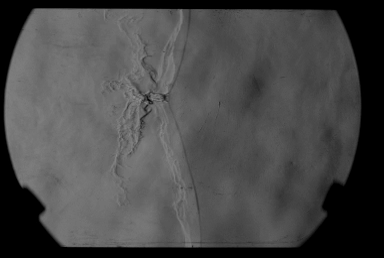} 
		%\caption{}
	\end{subfigure}
	\begin{subfigure}[c]{0.32\textwidth}
		\includegraphics[width=\linewidth]{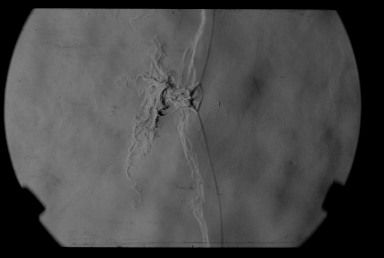} 
		%\caption{}
	\end{subfigure}
	\begin{subfigure}[c]{0.32\textwidth}
		\includegraphics[width=\linewidth]{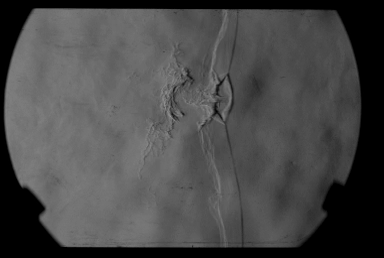} 
		%\caption{}
	\end{subfigure}
	
	\begin{subfigure}[c]{0.32\textwidth}
		\includegraphics[width=\linewidth]{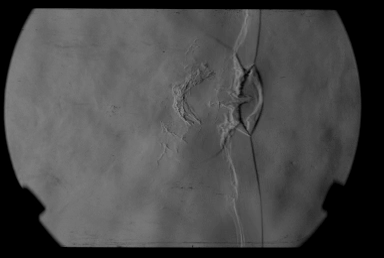} 
		%\caption{}
	\end{subfigure}
	\begin{subfigure}[c]{0.32\textwidth}
		\includegraphics[width=\linewidth]{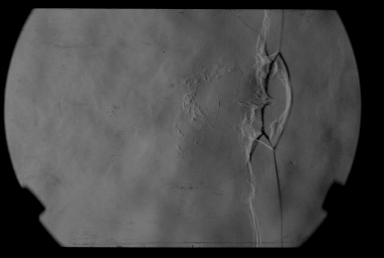} 
		%\caption{}
	\end{subfigure}
	\begin{subfigure}[c]{0.32\textwidth}
		\includegraphics[width=\linewidth]{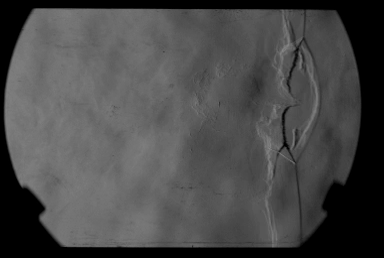} 
		%\caption{}
	\end{subfigure}
	
	\begin{subfigure}[c]{0.32\textwidth}
		\includegraphics[width=\linewidth]{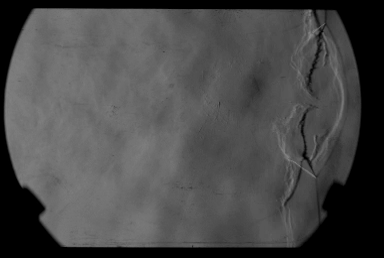} 
		%\caption{}
	\end{subfigure}
	\begin{subfigure}[c]{0.32\textwidth}
		\includegraphics[width=\linewidth]{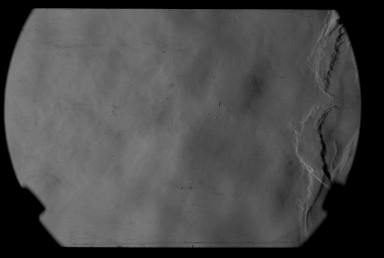} 
		%\caption{}
	\end{subfigure}
	\begin{subfigure}[c]{0.32\textwidth}
		\includegraphics[width=\linewidth]{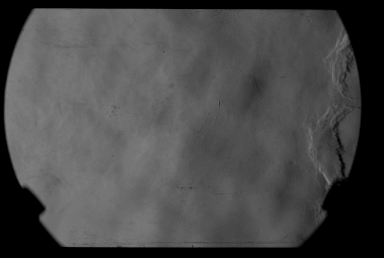} 
		%\caption{}
	\end{subfigure}
	\caption{Schlieren visualization of experiment 0\_CH4 performed in CH$_4$ + 2O$_2$. Taken from \cite{Maxwell2017}.}
	\label{fig:MethaneExpSummary}
\end{figure}

Figures \ref{fig:HydrogenExpSummary} and \ref{fig:MethaneExpSummary} show two typical high speed visualization sequences of the detonation structure seen in 2H$_2$ + O$_2$ + 7Ar and CH$_4$ + 2O$_2$, respectively, with the detonation propagating from left-to-right. The detonation wave seen in Figure \ref{fig:HydrogenExpSummary} begins near the end of a cell, with two Mach shocks near each of the walls and an incident shock in the centre, in black. The reaction waves, shown in white, follow the lead shocks, however the distance between the incident shock and the reaction wave is noticeably larger than the distance between the Mach shock and the reaction wave. Proceeding through the following two frames, one notices that the distance between the lead shock and the reaction wave increases. It is interesting to note the effect of the asymmetry on the transverse reaction waves, the topmost being closely coupled to its transverse shock wave whereas the bottom lags behind quite significantly.  The second row of frames sees the beginning of the next cell, starting at collision of the triple points. The top and bottom shocks now become the incident shocks, being weaker than the very strong Mach shock formed in the middle of the frame. The reaction wave in the centre follows the Mach shock very closely after the triple point collision, however after around nine frames the induction zone length begins to quickly increase. Similarly, this length will continuously increase throughout the remainder of the sequence. One can see the Mach shock runs ahead of the incident shocks in the subsequent frames due to the higher velocity associated with the high strength of this wave compared to the incident shock and the triple points.  The subsequent collision of the triple points occurs with the walls, out of the field of view. 

The evolution of the detonation propagating through the methane-oxygen mixture, shown in Figure \ref{fig:MethaneExpSummary}, looks similar to the previous mixture as many key features are shared between these detonations. The sequence begins near the end of a cell, with an incident shock in the centre of the cell bounded by two triple points which propagate towards one another. This incident shock is followed by a reaction zone separated by an increasing induction zone. The first qualitative difference appears in the reaction zone itself, which is more corrugated than the relatively featureless reaction zone seen in the previous mixture. This is similar to the lead shocks, which are again more corrugated than in the first mixture. Upon the collision of the triple points in the third row of frames, a pocket of unburnt gas in the post-shock region detaches from the shock and continues burning through turbulent combustion further behind the lead shock and its coupled reaction wave.

\section{Measurement of the Shock Dynamics}
\label{ch:ShockDynamicsMeas}
From the sequence of Schlieren images taken from the experiments, one can measure the position of the lead shock and its curvature. Measurement points are placed on the lead shock at its intersection with the top and bottom walls and along the cell's centreline, taken as halfway between the triple points. The velocity and acceleration of the lead shock is calculated using a centred difference scheme between frames. 

\begin{figure}
	\centering
	\begin{subfigure}[c]{0.32\textwidth}
		\includegraphics[width=\linewidth]{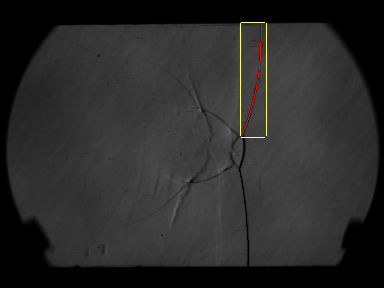} 
	\caption{}
	\end{subfigure}
	\begin{subfigure}[c]{0.32\textwidth}
		\includegraphics[width=\linewidth]{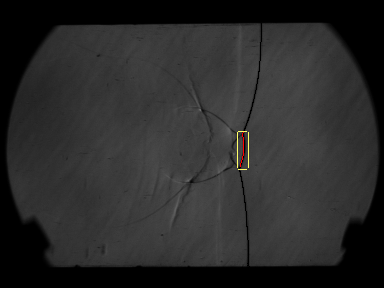} 
	\caption{}
	\end{subfigure}
	\begin{subfigure}[c]{0.32\textwidth}
		\includegraphics[width=\linewidth]{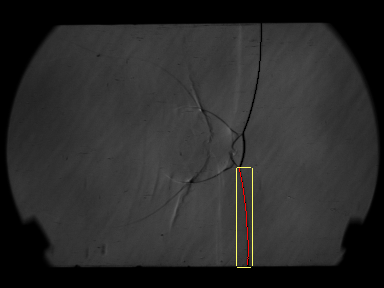} 
	\caption{}
	\end{subfigure}
	\caption{Measurement radius of curvature of the lead shock, for an experiment performed in argon-diluted hydrogen. The yellow boxes represent the manually measured regions of interest containing the lead shock, with the measured lead shock overlaid in red on the Schlieren frame.}
	\label{fig:RadiusMeasurement}
\end{figure}

In addition to the velocity of the lead shock, measurements of the curvature of the lead shock at the top wall, centreline, and bottom wall were also taken. A box enclosing the lead shock was measured at the walls and centreline. This interrogation window spans from the wall to the nearest triple point, or from one triple point to the other for the centremost shock. Once transformed into a series of points, an arc of circle was fit over these points using a least-square fitting method with a Levenberg-Marquardt algorithm. An example of such a measurement is shown in Figure \ref{fig:RadiusMeasurement}, which properly captures the entirety of the lead shock's position along its arc.

The top wall, centreline, and bottom wall measurements are separated into measurement series based on the triple point collision at those locations. Referring to Figure \ref{fig:HydrogenExpSummary}, the centreline has a series containing the first 4 frames (pre-collision), and a second series containing the remainder of the frames, whereas the top and bottom walls are comprised of a single series each (no data is separated by the triple point collisions as these occurred out-of-frame in this experiment). 

The pixelation of the wave in the images is found to give a measurement uncertainty of approximately 140 m/s for velocity measurements. The curvature uncertainty is obtained from fitting the arc of circle over the lead shock.

\subsection{Velocity and Curvature Progression}
\label{ch:VKFitting}
It was necessary to combine the separate series of measurements in such a way that an entire cell cycle can be plotted. As seen in Figures \ref{fig:HydrogenExpSummary} and \ref{fig:MethaneExpSummary}, the first visualized frame does not coincide with the beginning of a new cell. To reconstruct the cellular cycle, multiple methods were studied. A first method attempted to place each frame based on the ratio of the distance between the triple points and the height of the channel. In this method, the cell would follow the progress of the Mach shock from the triple point collision until the collision with the wall, followed by the progress of the incident shock in the second half of the cell. A second method studied the bottom- and top-half of the channel separately, and subsequently averaged the results, while following the reconstruction method of the first method. 

\begin{figure}
	\begin{subfigure}[c]{0.48\textwidth}
		\includegraphics[width=\linewidth]{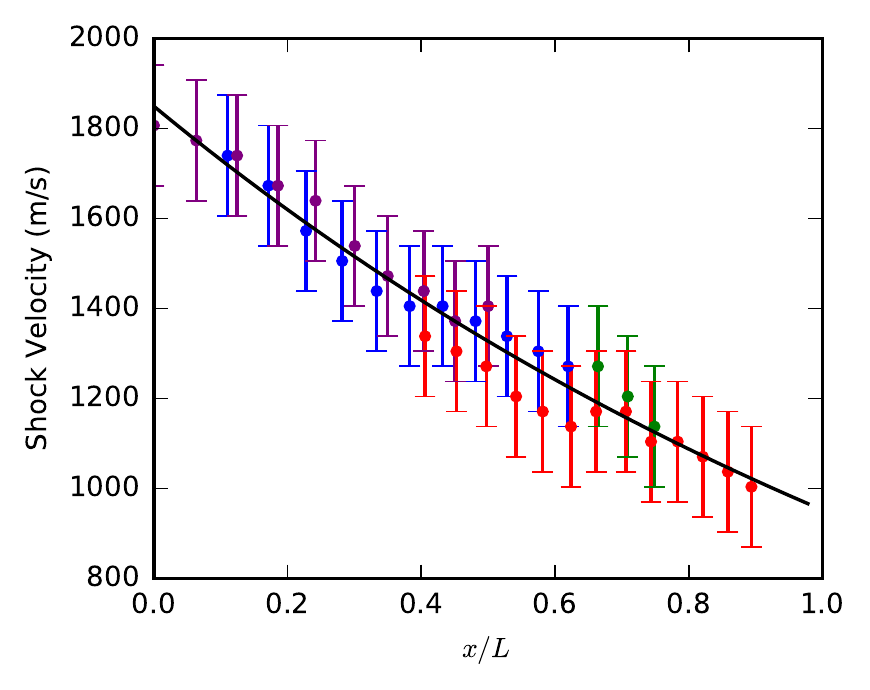} 
		\caption{}
	\end{subfigure}
	\begin{subfigure}[c]{0.48\textwidth}
		\includegraphics[width=\linewidth]{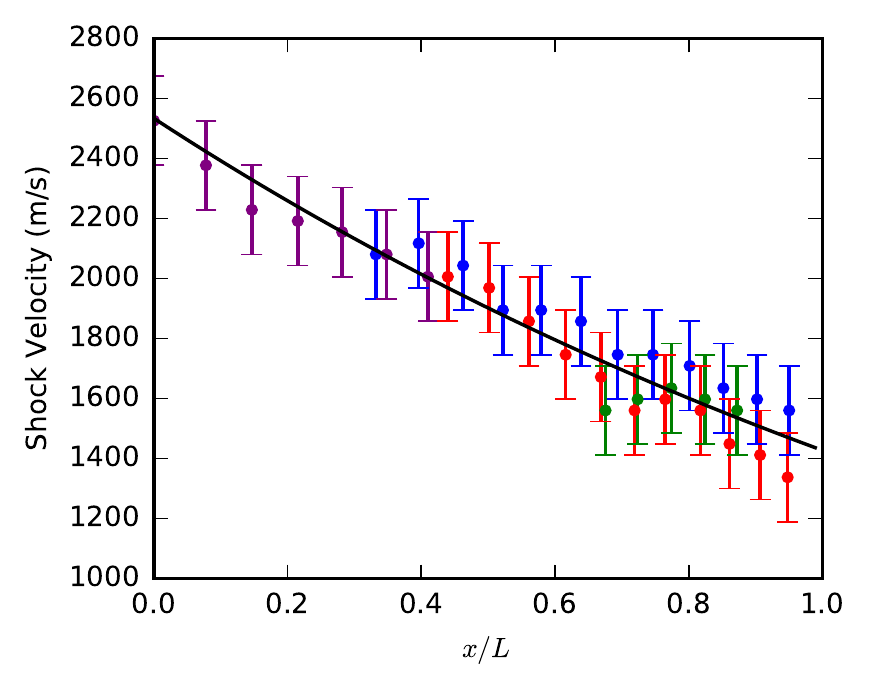} 
		\caption{}
	\end{subfigure}
	\caption{The shock speed evolution over a detonation cell, with the exponential curve-fit traced in black. For experiments a) 13\_H2 and b) 0\_CH4. The colour of each point represents their measurement series - blue : top wall, red : bottom wall, green : centreline pre-collision, purple : centreline post-collision.}
	\label{fig:VelocityMeasurementPlot}
\end{figure}

The method presented below consists of reconstructing the detonation cell by shifting the position of each data series in order to obtain a continuous speed in the cell cycle. This method is possible because the beginning and end of each series of measurements share significant regions of overlapping speeds. Figure \ref{fig:VelocityMeasurementPlot} shows the speed evolution of a reconstructed cell, with each series plotted in a different colour. The error associated with this reconstruction method also agrees with the exponential trendline, and accounts for the reconstruction error. In Figures \ref{fig:HydrogenExpSummary} and \ref{fig:MethaneExpSummary}, the four measurement series are attributed to the lead shock evolution at the top wall (middle of cell), bottom wall (middle of cell), centreline before the triple point collision (end of cell), and centreline after the triple point collision (beginning of cell). Experiments 5\_H2 and 11\_H2 has a collision of a triple point with the top and bottom walls, thus separating these measurement series at the collision is necessary.

\begin{figure}
	\begin{subfigure}[c]{0.48\textwidth}
		\includegraphics[width=\linewidth]{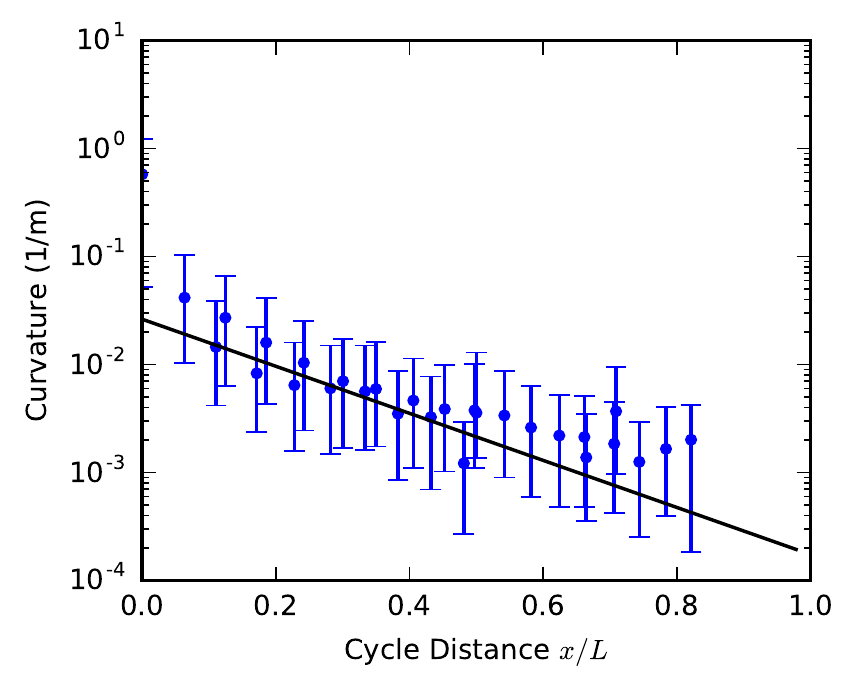} 
		\caption{}
	\end{subfigure}
	\begin{subfigure}[c]{0.48\textwidth}
		\includegraphics[width=\linewidth]{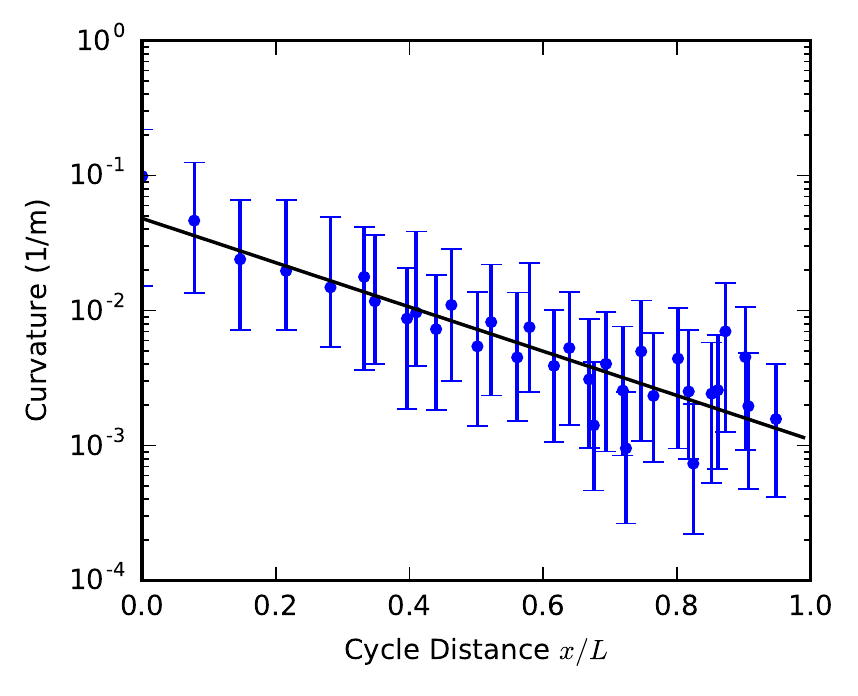} 
		\caption{}
	\end{subfigure}
	\caption{The evolution of the logarithm of curvature over a detonation cell, with the exponential curve-fit traced in black. For experiments a) 13\_H2 and b) 0\_CH4. One notices the large early-cell curvature value of 0.6m at x=0 in a), which was not included in the fit.}
	\label{fig:CurvatureMeasurementPlot}
\end{figure}

Once the cell is reconstructed, one can curve-fit the progression of the speed and the curvature over the cell distance. These fits are shown in Figures \ref{fig:VelocityMeasurementPlot} and \ref{fig:CurvatureMeasurementPlot}. The exponential functions 
\begin{equation}
	D = D_0 e^{-a x}, \label{eq:VelocityFit}
\end{equation}
and 
\begin{equation}
\kappa = \kappa_0 e^{-b x},
\end{equation}
are chosen for these curve-fits as they are found to offer better fits than 3rd order polynomial and power law fits. The choice of using an exponential decay is also supported by the findings of \cite{JACKSON2018}, where an exponential speed decay is found. One notices the generally good agreement between the fitted curves and the measured data, however a linear decay would have fit the data equally well. The parameters of the curve fits are shown in tables \ref{tab:VKCurveFit} along with the R-squared value associated to each fit. When fitting the evolution of curvature over the cell, the very large curvature values at the beginning of the cell are removed as outliers. This is in part to the difficulty of measuring a radius on the order of millimetres from the pixelated frames reducing the meaningfulness of such measurement points, and in part due to these points being outliers due to their extremely large value which is frequently on the order of meters. In comparison to this, the remainder of the curvature values very rarely reach an order of $10^{-1}$ m. The fitted curve under-predicts the curvature at the beginning of the cell regardless of the inclusion of these points in the fit, therefore an accurate reconstruction of the evolution of curvature over the remainder of the cell is preferred.

\begin{table}
	\centering
	\caption{Curvature and shock speed curve fitting parameters.}
	\vspace{-0.15cm}
	\label{tab:VKCurveFit}
	\begin{tabular}{c c c c c c c}
		\hline \hline
		Experiment  & $D_0$ (m/s) & $a$ & $R^2_D$ & $\kappa_0$ (1/m) & $b$ & $R^2_\kappa$ \\
		\hline
		0\_H2 & $1783$ & $1.8$ & $0.92$ & $0.02$ & $9.1$ & $0.95$ \\
		5\_H2 & $1902$ & $1.9$ & $0.93$ & $0.029$ & $10.9$ & $0.92$ \\
		11\_H2 & $1812$ & $2.1$ & $0.96$ & $0.027$ & $14.3$ & $0.68$ \\
		13\_H2 & $1849$ & $1.8$ & $0.96$ & $0.026$ & $13.6$ & $0.79$ \\
		18\_H2 & $1806$ & $1.7$ & $0.97$ & $0.023$ & $9.9$ & $0.80$ \\
		19\_H2 & $1934$ & $3.0$ & $0.95$ & $0.039$ & $14.6$ & $0.68$ \\
		0\_CH4 & $2534$ & $1.5$ & $0.93$ & $0.048$ & $10.1$ & $0.86$ \\
		\hline
	\end{tabular}
\end{table}

\begin{figure}
	\centering
	\includegraphics[width=0.8\textwidth]{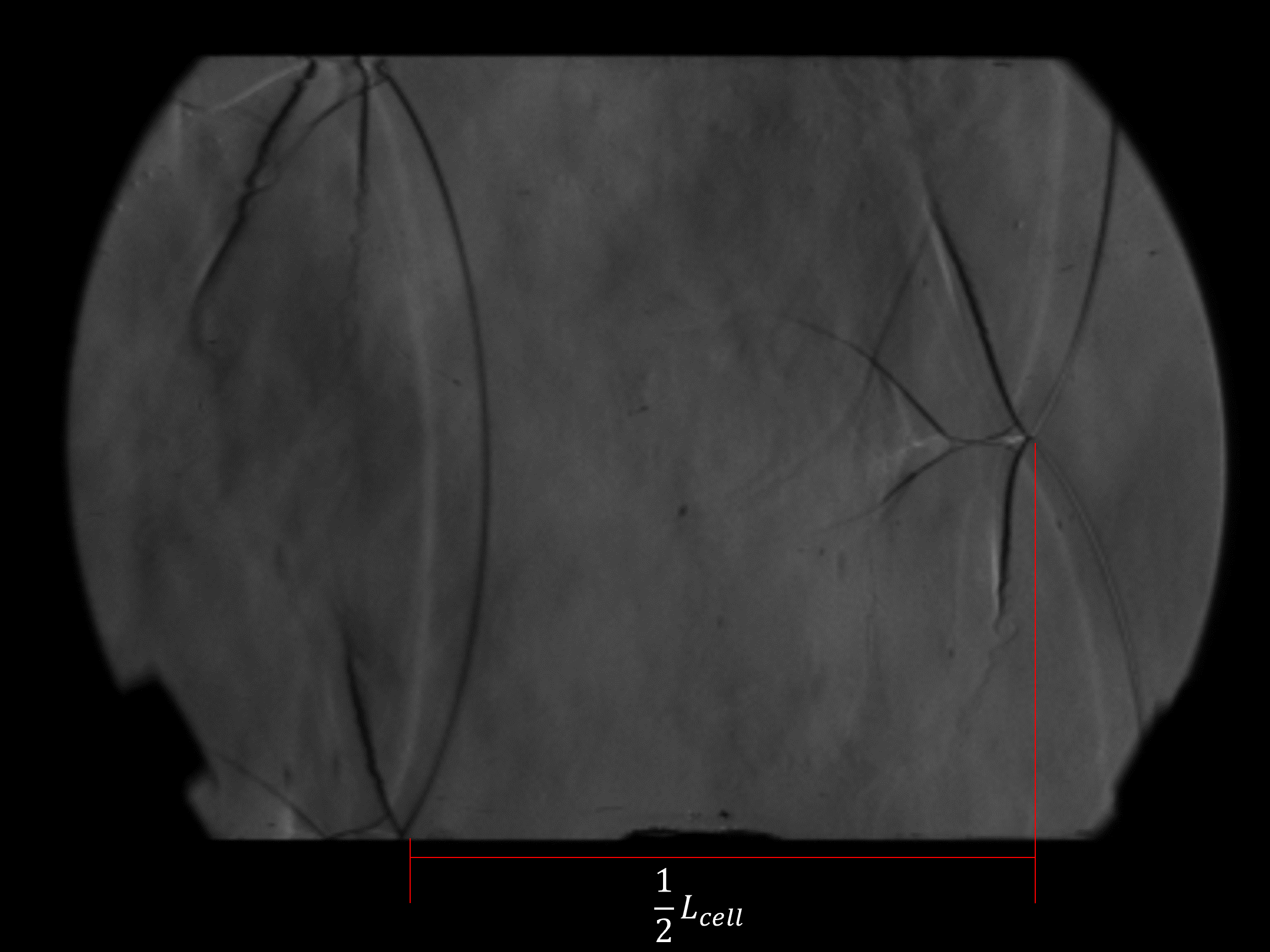}
	\caption{Example of the measurement of the cell length from experimental frames.}
	\label{fig:CycleLengthMeasurement}
\end{figure}

\subsection{Cell Length}
\label{ch:CellLength}
\begin{figure}
	\begin{subfigure}[c]{0.48\textwidth}
		\includegraphics[width=\linewidth]{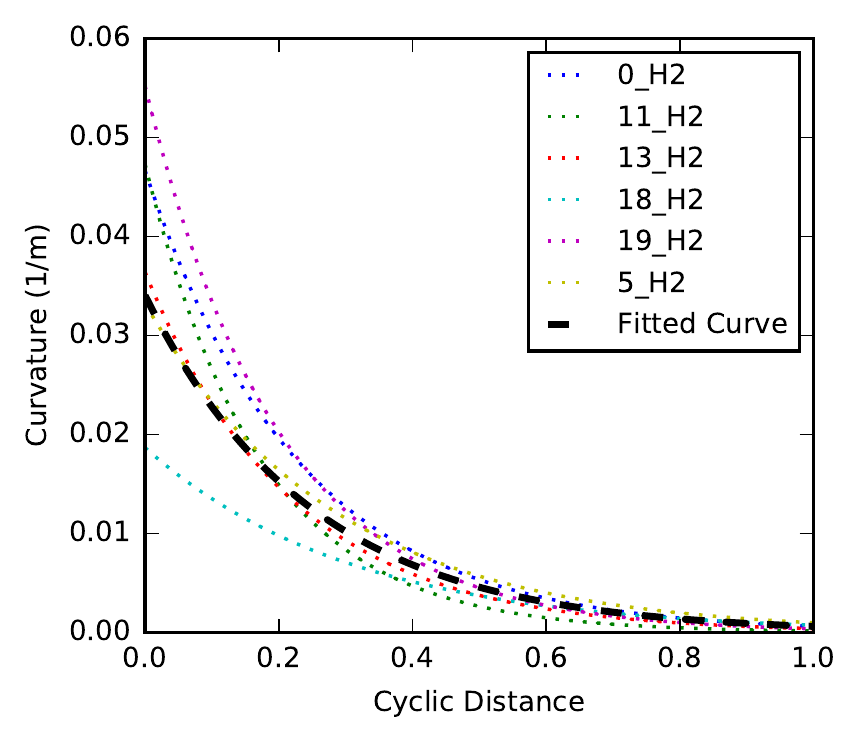} 
		\caption{}
	\end{subfigure}
	\begin{subfigure}[c]{0.48\textwidth}
		\includegraphics[width=\linewidth]{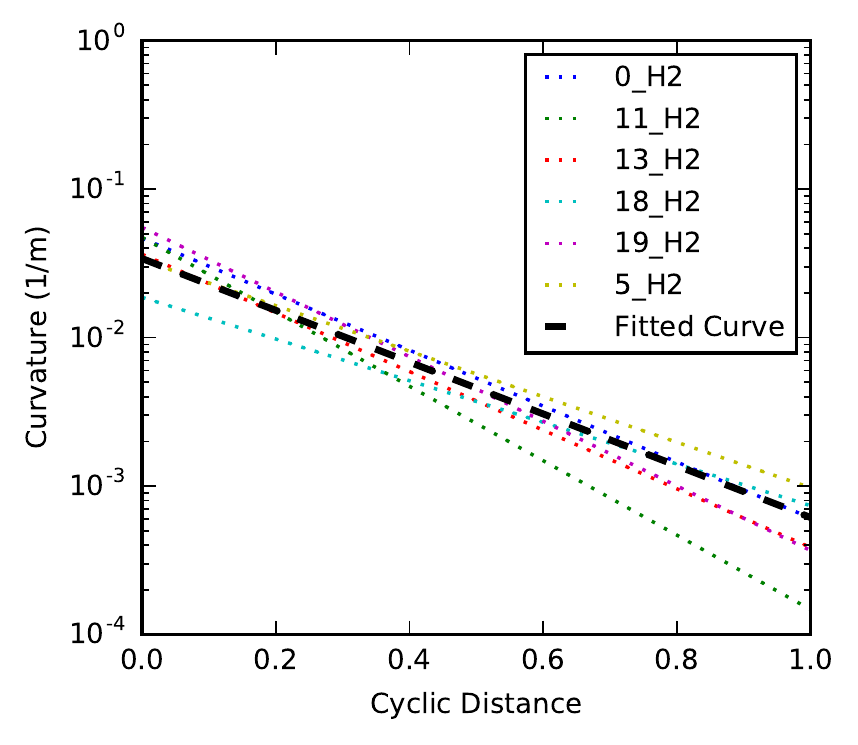} 
		\caption{}
	\end{subfigure}
	\caption{The curve-fit evolution of the curvature plotted over the cell length, for the experiments performed in argon-diluted hydrogen-oxygen. A curve fit representative of the average evolution of curvature is also presented.}
	\label{fig:CollapsedCurvature}
\end{figure}

\begin{figure}
	\begin{subfigure}[c]{0.48\textwidth}
		\includegraphics[width=\linewidth]{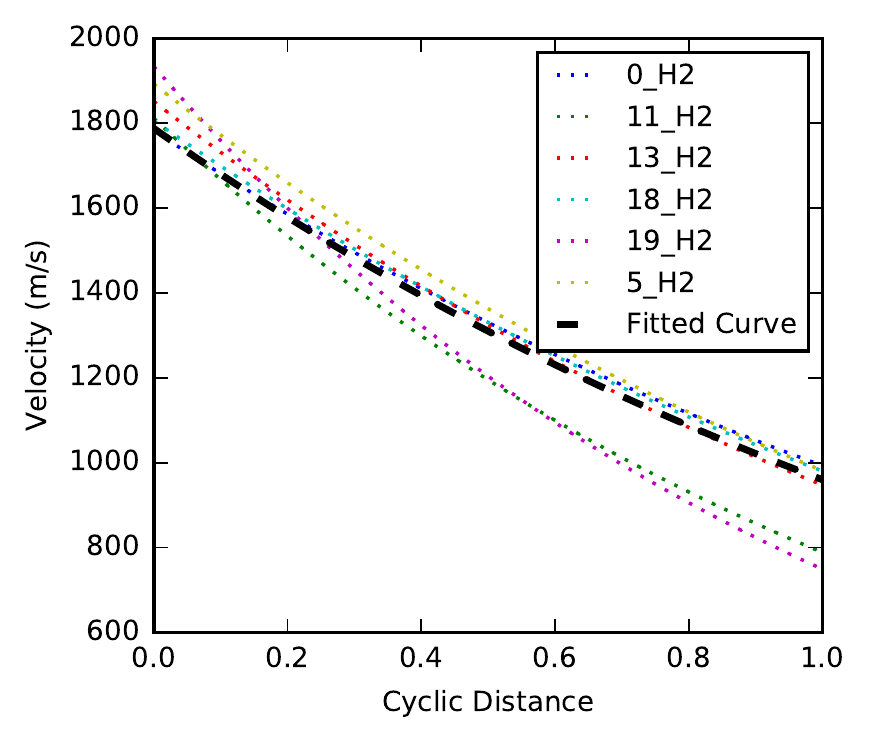} 
		\caption{}
	\end{subfigure}
	\begin{subfigure}[c]{0.48\textwidth}
		\includegraphics[width=\linewidth]{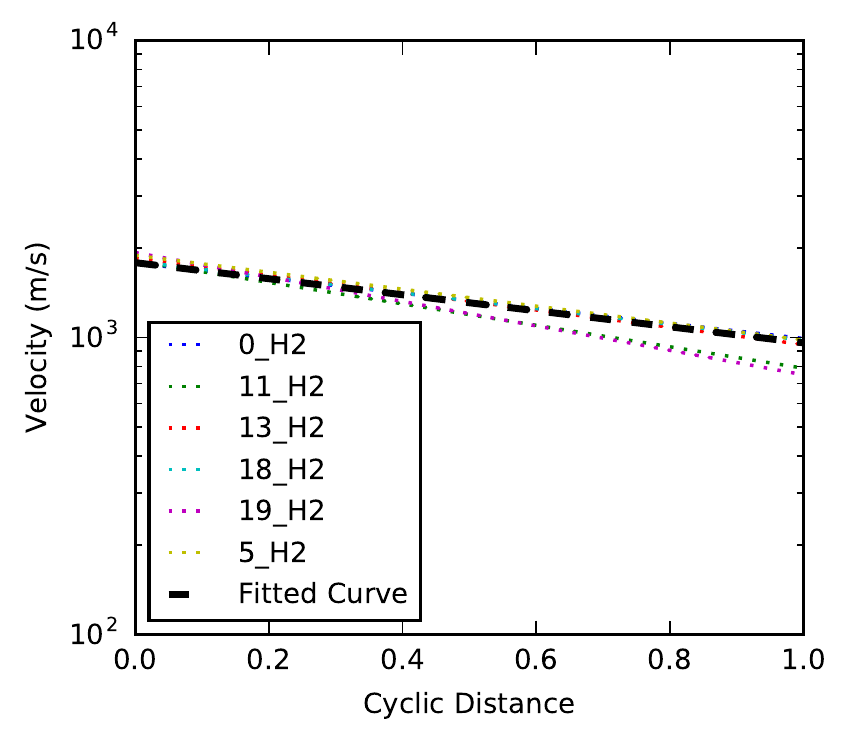} 
		\caption{}
	\end{subfigure}
	\caption{The curve-fit evolution of the lead shock velocity plotted over the cell length, for the experiments performed in argon-diluted hydrogen-oxygen. A curve fit representative of the average evolution of the lead shock is also presented.}
	\label{fig:CollapsedVelocity}
\end{figure}

By the assumption of lateral symmetry in detonation cells, the cycle length is determined by measuring the half-length of the cell from the collision of the triple points at the centreline and at the wall, as shown in Figure \ref{fig:CycleLengthMeasurement}. To compare the cells across all of the experiments, the spatial distances are normalized using the measured cyclic length $L_{cell}$ of each experiment. Once normalized, the shock speed and curvature can be plotted over a cycle distance ranging from 0 at the beginning of the cycle to 1 at its end, seen in figures \ref{fig:CollapsedVelocity} and \ref{fig:CollapsedCurvature}. These figures contain a comparison of all the experiments performed in the argon-diluted hydrogen-oxygen mixture, and plots a curve representative of all the data. This average curve is found to closely follow the evolution of the experiment 13\_H2. 

\subsection{Average Propagation Speed and Velocity Deficit}
The average propagation velocity of the detonation cell can be calculated either from the curve-fit velocity evolution presented in section \ref{ch:VKFitting}, or by directly measuring the average velocity from the visualized frames. The first method consists of finding the average of the exponential fit presented in section \ref{ch:VKFitting}. This is found from the expression
\begin{equation}
\bar{D} = \frac{L_{cell}-0}{t_{cell}-0}, 
\end{equation}
with $t_{cell}$ found by integrating the velocity
\begin{equation}
\frac{dx}{dt} = D(x) = D_0 e^{-a x},
\end{equation}
which can be recast as
\begin{equation}
\int_0^{L_{cell}} \frac{dx}{D_0 e^{-ax}} = \int_0^{t_{cell}} dt.
\end{equation}
Solving this integral yields
\begin{equation}
t_{cell} = \frac{1}{a}\left[\frac{1}{D(L_{cell})} - \frac{1}{D_0}\right].
\end{equation}
This cell time gives an expression for the average speed of
\begin{equation}
\bar{D} = a L_{cell} \frac{D_L D_0}{D_0-D_L},
\end{equation}
with $D_L = D(L_{cell})$. For the experiment 0\_H2, an average velocity of $1310$ m/s is found from $D_0=1783$ m/s, $a=1.8$, and $L_{cell}=0.325$ m/s taken from Table \ref{tab:VKCurveFit} and $D_L = 993$ m/s calculated from (\ref{eq:VelocityFit}). 

Another method to verify the accuracy of the fit curves is available by measuring an average propagation velocity directly from the experimental frames. For this measurement, an early frame near a triple point collision is chosen for an unambiguous measurement of the lead shock location. If a frame is chosen containing a wave slightly before the triple point collision, the second frame chosen will also contain a wave slightly before a triple point collision. The height of the incident shock/Mach shock should be similar in both frames. The location of the wave is measured for the top and bottom wall independently, as well as the timespan between the frames associated with each measurement. An average propagation velocity for the top and bottom half of the cell is calculated as 
\begin{equation}
\bar{D} = \frac{x_2 - x_1}{\Delta t}, 
\end{equation} 
ans subsequently averaged to obtain the average propagation velocity of the overall wave. This measurement of velocity is also shown in Table \ref{tab:AvgSpeeds}, with an average velocity of 1280 m/s measured for experiment 0\_H2. 

The average propagation velocity found using both methods are within 138 m/s of one another for every experiment, which is the uncertainty associated with measurements of velocity. Xiao and co-workers measured the average velocity using a time-of-arrival method with wall-mounted pressure transducers separated by a distance of over 1 meter, reporting an average velocity deficit of 0.83$D_{CJ}$ for 0\_H2 \citep{Xiao2020}. Maxwell reports a velocity deficit of 0.83$D_{CJ}$ for experiment 0\_CH$_4$ \citep{Maxwell2017}. Both these reported velocity deficits are in excellent agreement with both measurements presented for calculating the velocity deficit, and further validates the method used to reconstruct the detonation cells presented in section \ref{ch:VKFitting}. 

\begin{table}
	\centering
	\caption{Detonation cell scales and average propagation velocities.}
	\vspace{-0.15cm}
	\label{tab:AvgSpeeds}
	\begin{tabular}{c c c c c c c}
		\hline \hline
		Experiment & $\bar{D}_{fit}$ (m/s) & $\bar{D}_{meas}$ (m/s) & $\frac{\bar{D}_{meas}}{D_{CJ}}_{fit}$ & $\frac{\bar{D}_{meas}}{D_{CJ}}_{meas}$ & $L_{cell}$ (m) & $t_{cell}$ (s) \\
		\hline
		0\_H2 & $1310$ & $1280$ & $0.82$ & $0.80$ & $0.325$ & $0.00025$ \\
		5\_H2 & $1330$ & $1410$ & $0.83$ & $0.88$ & $0.357$ & $0.00025$ \\
		11\_H2 & $1160$ & $1320$ & $0.72$ & $0.82$ & $0.396$ & $0.00030$ \\
		13\_H2 & $1300$ & $1400$ & $0.81$ & $0.87$ & $0.368$ & $0.00026$ \\
		18\_H2 & $1310$ & $1380$ & $0.82$ & $0.86$ & $0.355$ & $0.00026$ \\
		19\_H2 & $1130$ & $1240$ & $0.71$ & $0.78$ & $0.331$ & $0.00027$ \\
		0\_CH4 & $1880$ & $1870$ & $0.84$ & $0.83$ & $0.374$ & $0.00020$ \\
		\hline
	\end{tabular}
\end{table}

\subsection{Shock Change Equation}
\label{ch:ShockChangeEquation}
\begin{figure}
	\begin{subfigure}[c]{0.48\textwidth}
		\includegraphics[width=\linewidth]{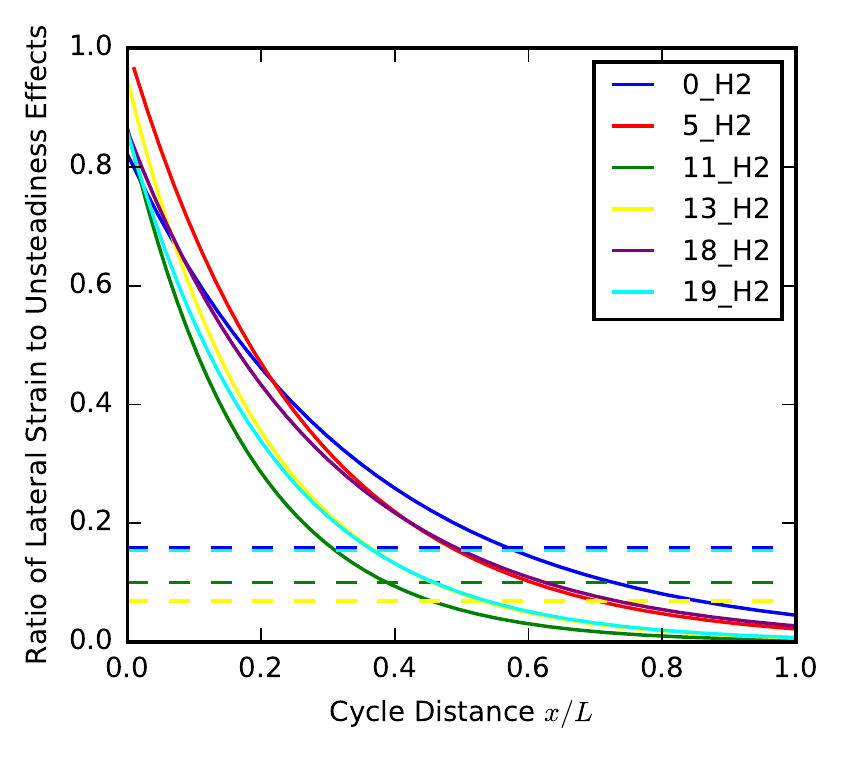} 
		\caption{}
	\end{subfigure}
	\begin{subfigure}[c]{0.48\textwidth}
		\includegraphics[width=\linewidth]{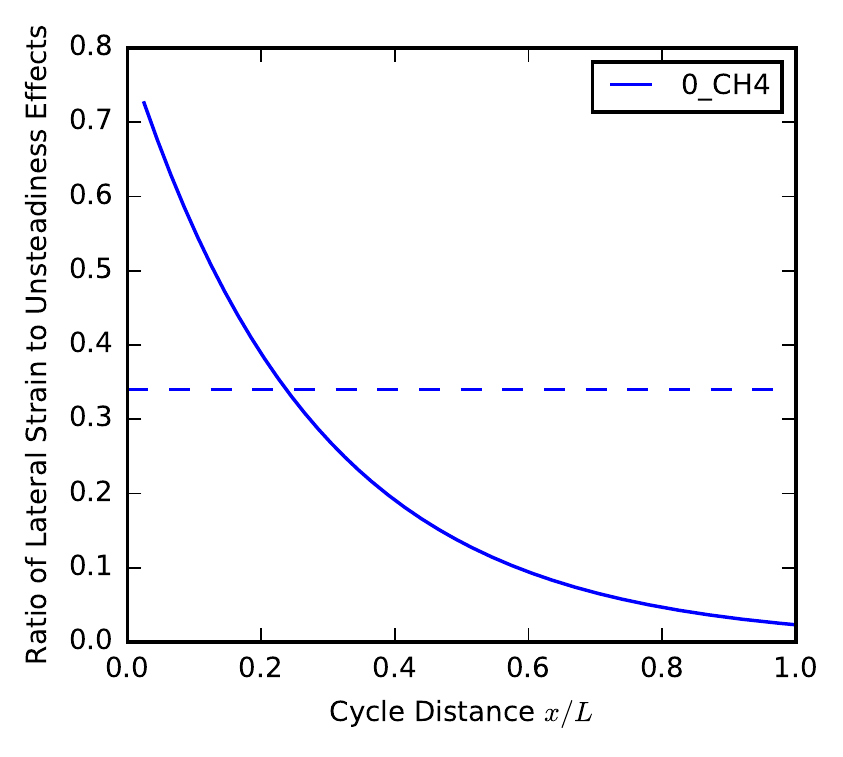} 
		\caption{}
	\end{subfigure}
	\caption{Ratio of lateral strain rate to unsteadiness terms, with the horizontal intersection with the critical ignition limit of each cycle as a dotted line of the same colour. For a) argon-diluted hydrogen-oxygen, and b) stoichiometric methane-oxygen.}
	\label{fig:RatioPlots}
\end{figure}

The shock change equation contains two competing terms, best seen under the assumption of a strong shock (\ref{eq:StrongShockChange}). As previously discussed in section \ref{ch:ShockDynamics}, the rate of lateral strain due to changing curvature and the shock unsteadiness terms are only a function of the initial state, and the velocity and curvature of the shock front. We can now compare them throughout the cycle. At the beginning of the cycle, the ratio of both terms begins on the order of 1, with the unsteadiness term being slightly larger than the rate of lateral strain of the lead shock. This is the case for both studied mixtures. The rate of lateral strain quickly decreases to become smaller by an order of magnitude at around half the cycle length as seen in Figure \ref{fig:RatioPlots}. Horizontal dotted lines are plotted to show the critical ratio of these terms at which quenching of the post-shock reaction occurs. The quenching of the post-shock reactions will be discussed in more details in section \ref{ch:ChemKinetics}. 

From Figure \ref{fig:RatioPlots}, one notices that quenching of the post-shock reaction occurs in both mixtures when the rate of lateral strain is an order of magnitude smaller than the shock unsteadiness, with the critical rate being a factor of 3 larger in the methane-oxygen mixture. This may be linked to the highly unstable mixture's higher sensitivity to fluctuations in temperature than the weakly unstable mixture. As quenching occurs in both mixtures when the magnitude of the lateral strain rate becomes much smaller than the unsteadiness, one can infer that the quenching mechanism is not solely a function of the decay of the strength of the lead shock. The change in lead shock curvature plays an important role in the ignition processes. However, as the shock flattens and the rate of lateral strain becomes negligible in the latter portion of the cycle, the curvature ceases to be an important effect in controlling the dynamics of the lead shock, which is then dominated by the lead shock's decay rate. 

\section{Ignition with Full Chemistry}
\label{ch:ChemKinetics}
We now study the evolution of the ignition process occurring in the post-shock state throughout a  detonation cell cycle. As described in section \ref{ch:CanteraDescription}, the initial state and the volumetric expansion rate of a Lagrangian particle crossing the lead shock at a given instant is calculated from the Rankine-Hugoniot shock jump equations and the shock change equations, respectively. The required measurements to describe the dynamics of the lead shock were presented in section \ref{ch:ShockDynamicsMeas}, i.e., the lead shock speed, shock decay rate, and its curvature. The detonation cell was reconstructed and lead shock dynamics were curve-fit in section \ref{ch:VKFitting}. This reconstruction of the detonation cell and fitting of the shock dynamics allows one to calculate the post-shock state and volumetric expansion for any particle path which crosses the lead shock. 

\begin{figure}
	\begin{subfigure}[c]{0.48\textwidth}
		\includegraphics[width=\linewidth]{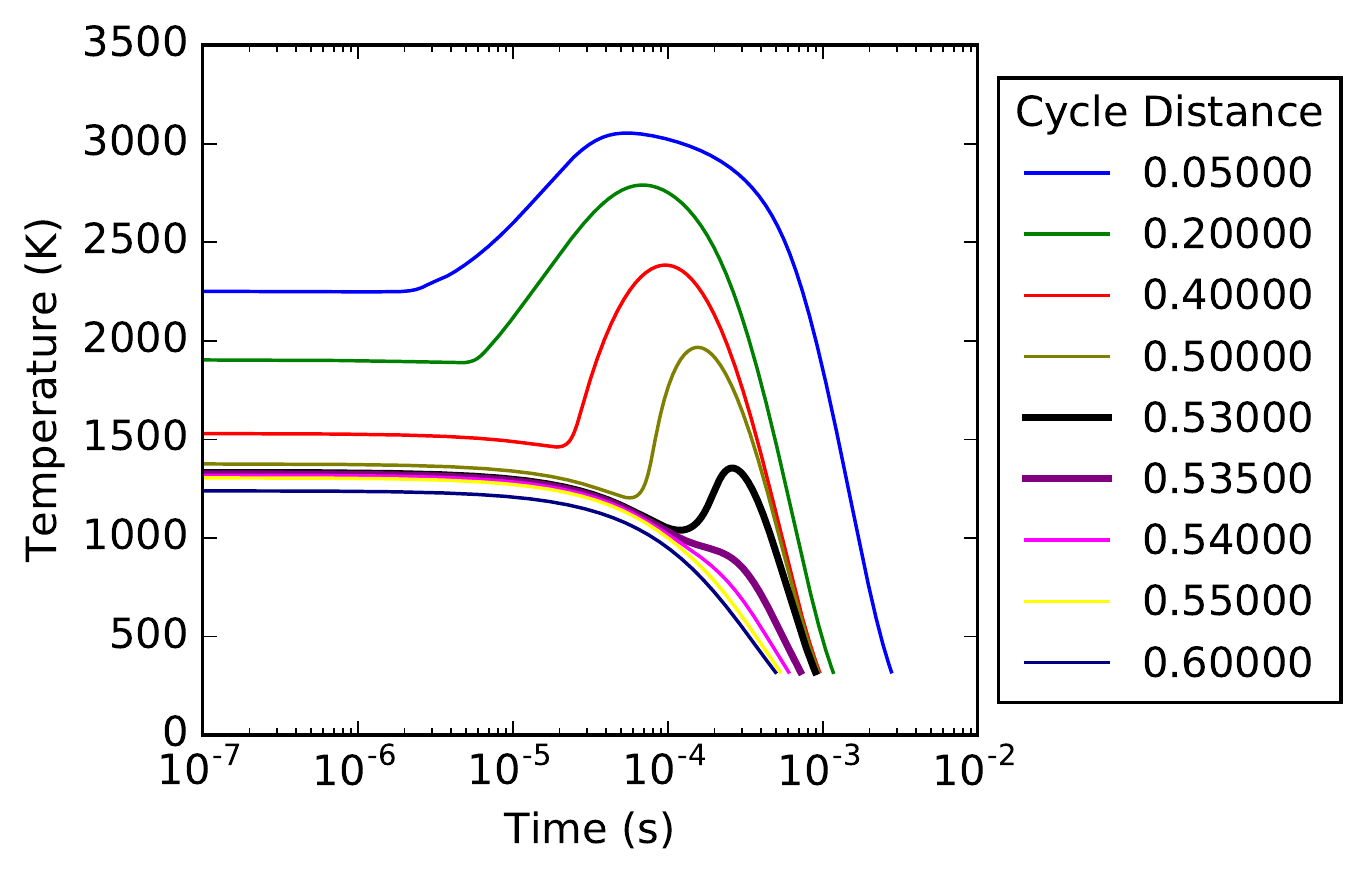} 
		\caption{}
	\end{subfigure}
	\begin{subfigure}[c]{0.48\textwidth}
		\includegraphics[width=\linewidth]{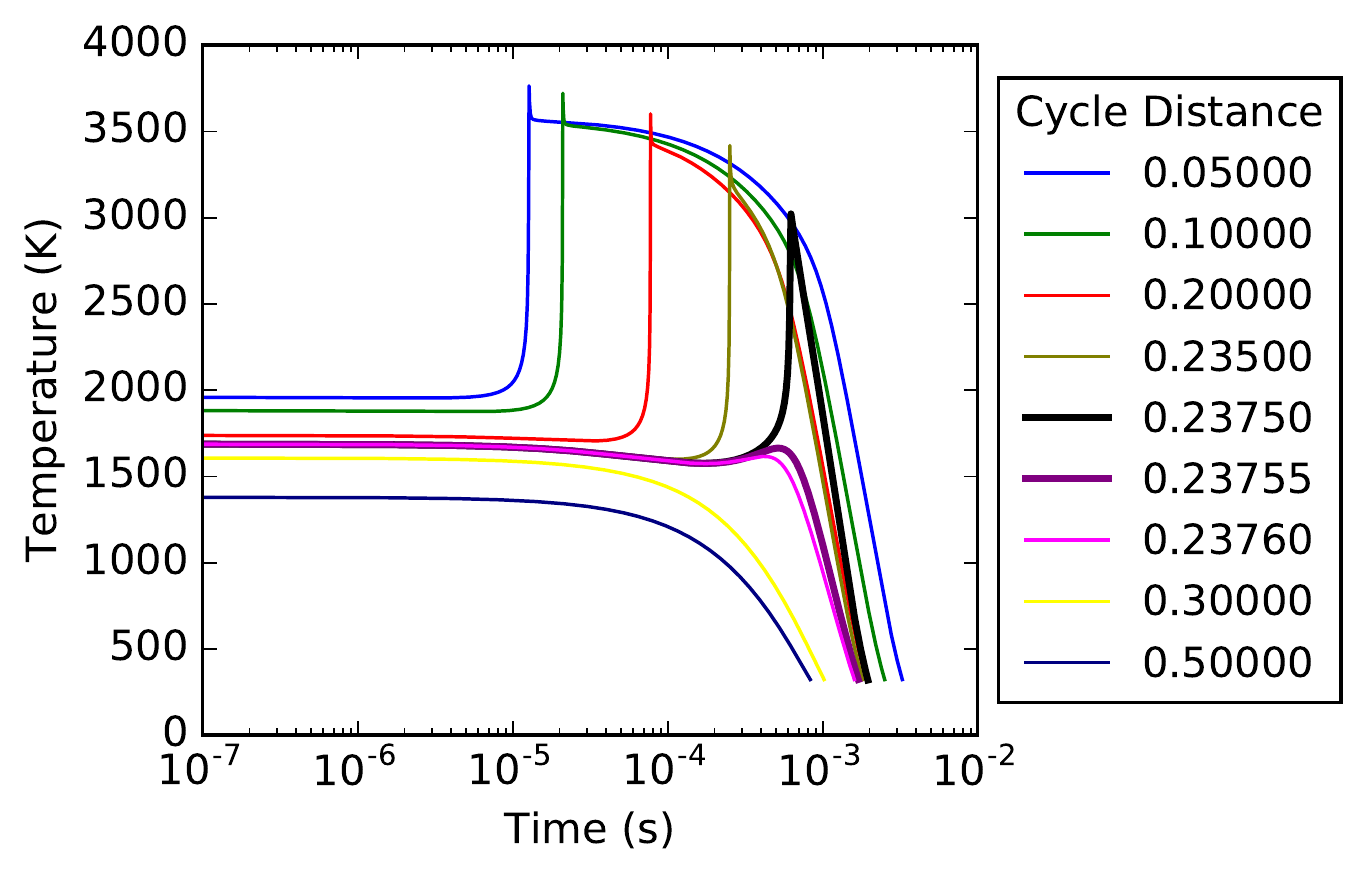} 
		\caption{}
	\end{subfigure}
	\caption{Plot of the evolution in time of the temperature in the post-shock state, for selected particle paths labelled at the location of crossing of the lead shock. Plotted experiments are 13\_H2 and 0\_CH4, for which the critical ignition limit is associated with the particle paths 0.5300$L_{cell}$ and 0.2375$L_{cell}$.}
	\label{fig:TemperatureEvolutionPlot}
\end{figure}

Figures \ref{fig:TemperatureEvolutionPlot} a) and b) show the evolution of temperature along a particle path as a function of time, for particles crossing the lead shock at increasing cell distances. Two general behaviours appear, which depend on the cell distance at which the particle path crosses the lead shock - quenching of the post-shock reactions occurring later in the cell and the occurrence of an ignition process occurring early in the cell.  These two behaviours are separated by a critical path, which is taken as the final particle path which sees the occurrence of ignition in the post-shock region. Studying the particle paths which cross the lead shock early in the cell cycle, one sees that the post-shock temperature slowly decreases throughout the the induction zone until a drastic increase in temperature occurs. This temperature increase is large enough to cause the thermal runaway associated with ignition events. At later times, once the energy has been released, the model ceases to be valid as the prescribed rate of expansion for the induction zone may depart significantly from its post shock value.   Nevertheless  the predicted temperature slowly decreases until the maximum integration time is reached or until the 323 K temperature floor of the combustor simulation is reached. This behaviour is similar to the evolution of temperature shown in Figure \ref{fig:CJPostShock} a), associated with a small non-zero volumetric expansion rate. As the volumetric expansion rate increases, the peak temperature decreases and time elapsed until the peak temperature is attained also increases. The rate at which the temperature decays prior to ignition and after the completion of the combustion process also increases with the post-shock expansion rate. 

When the post-shock expansion is increased past a critical threshold, as in the largest volumetric expansion rate plotted in Figure \ref{fig:CJPostShock}, the second behaviour of quenched reactions is seen. This begins with the particle path plotted in bold purple in Figures \ref{fig:TemperatureEvolutionPlot} a) and b), and continues for the remainder of the cell. Particle paths with quenched post-shock reactions see a decreasing temperature throughout the post-shock region due to the absence of the thermal runaway process which occurs during ignition. The rate of energy release is unable to overcome the expansion cooling imposed on the particle in the post-shock state. This decrease in temperature continues until the 323 K temperature floor of the combustor simulation is reached. 

The critical path, plotted as a bold black curve in Figure \ref{fig:TemperatureEvolutionPlot}, represents the last particle path along which a Lagrangian particle crossing the lead shock will ignite. This critical path shares the characteristic rise in temperature occurring during ignition with all previous particle paths in the cell. Due to the substantial effect of expansion cooling compared to previous particle paths which were subjected to smaller volumetric expansion rates in the post-shock region, this critical path is associated to the longest ignition delay of any particle path crossing the lead shock. The change in behaviour occurs very abruptly after this critical path, with a particle path crossing the lead shock at a distance of $0.005 L_{cell}$ after the critical path clearly having quenched post-shock reactions, as seen in Figure \ref{fig:TemperatureEvolutionPlot}.

\begin{figure}
	\begin{subfigure}[c]{0.48\textwidth}
		\includegraphics[width=\linewidth]{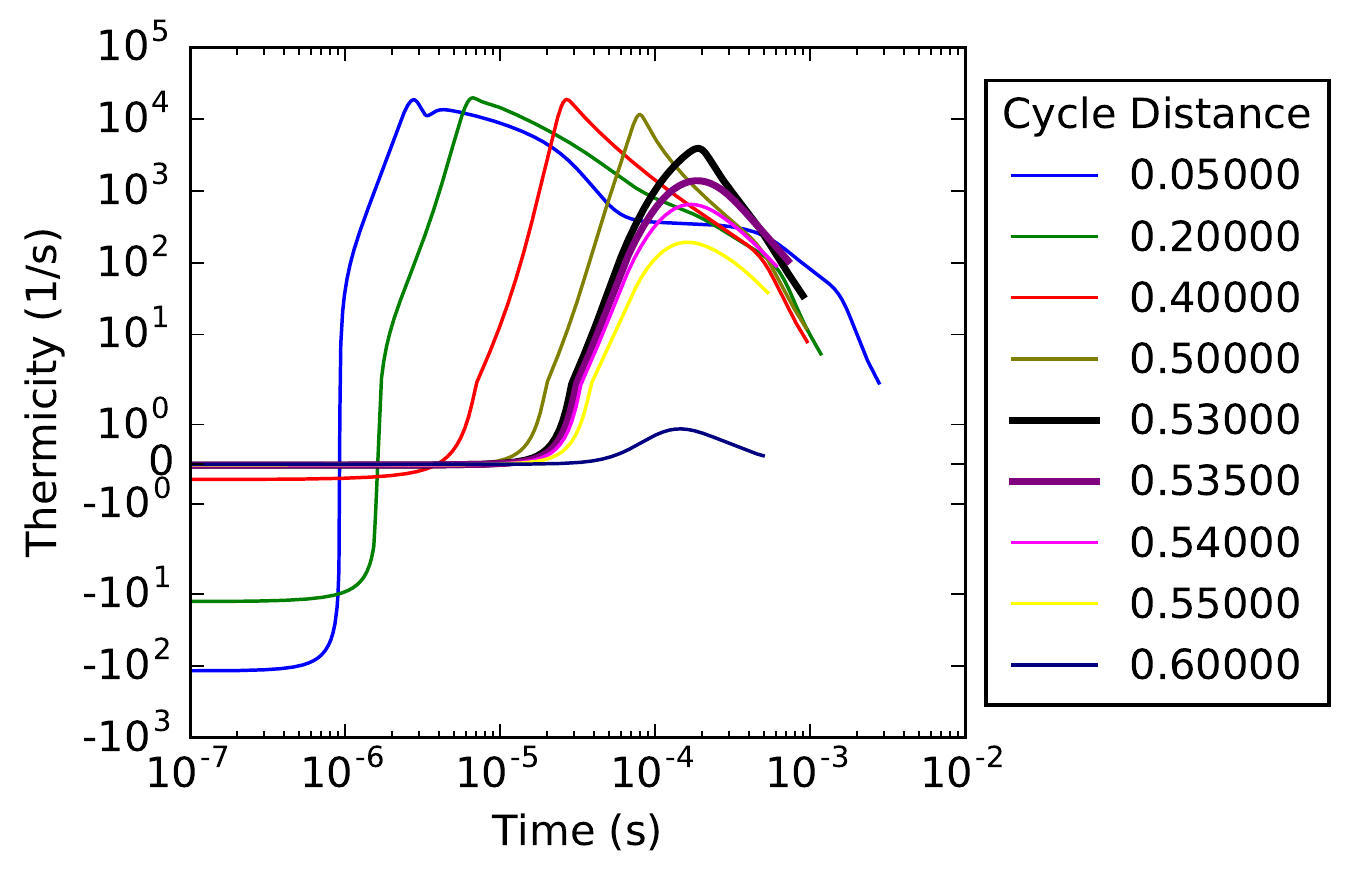} 
		\caption{}
	\end{subfigure}
	\begin{subfigure}[c]{0.48\textwidth}
		\includegraphics[width=\linewidth]{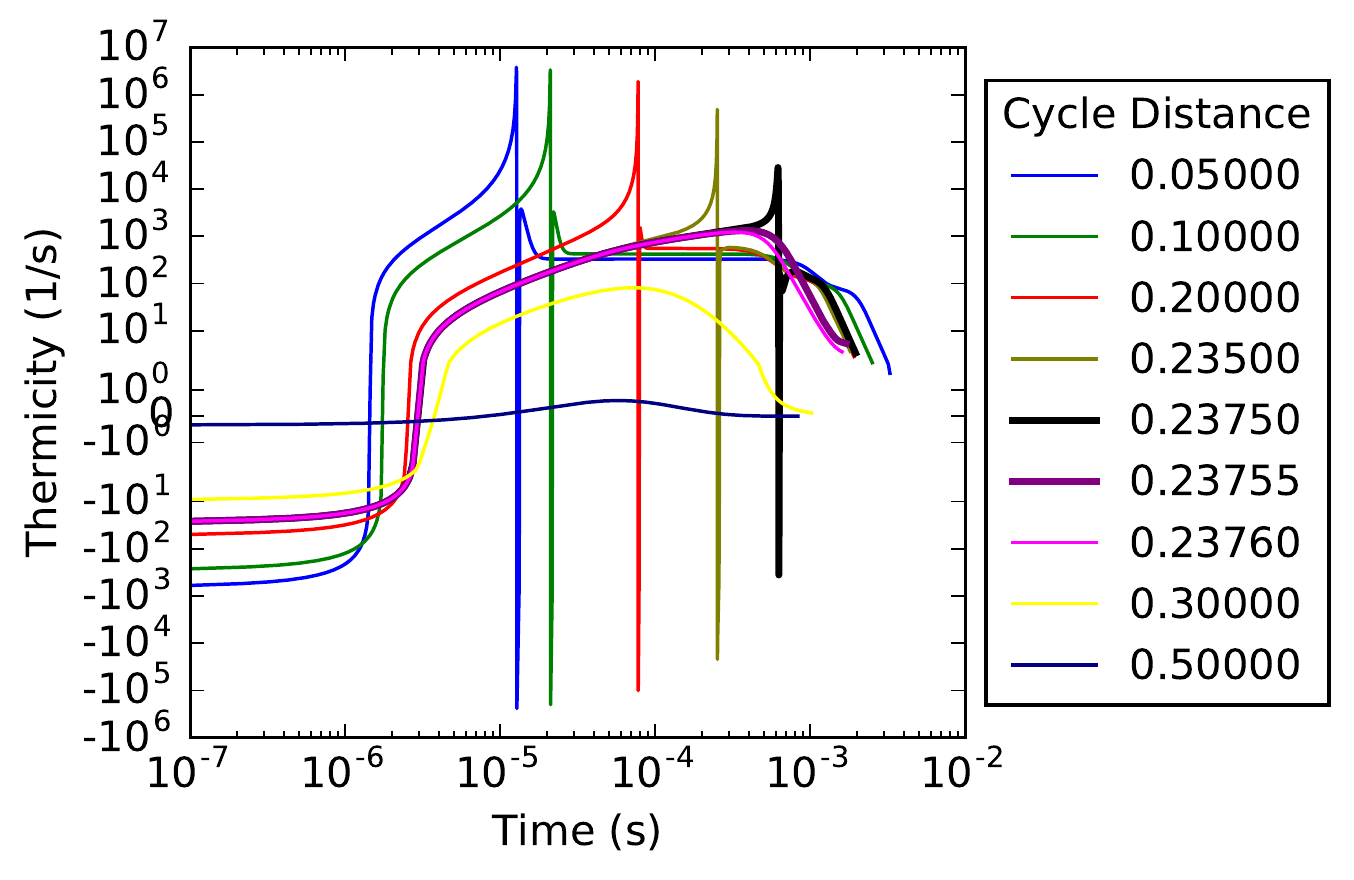} 
		\caption{}
	\end{subfigure}
	\caption{Plot of the evolution in time of the thermicity in the post-shock state, for selected particle paths labelled at the location of crossing of the lead shock. Plotted experiments are 13\_H2 and 0\_CH4, for which the critical ignition limit is associated with the particle paths 0.5300$L_{cell}$ and 0.2375$L_{cell}$.}
	\label{fig:ThermicityEvolutionPlot}
\end{figure}

Similar to the evolution of temperature, the evolution of the rate of heat release, or thermicity, over a detonation cell can also give insight as to the ignition behaviour along a particle path. Figure \ref{fig:ThermicityEvolutionPlot} shows the evolution of thermicity, as a function of time along the same particle paths as plotted in Figure \ref{fig:TemperatureEvolutionPlot}. The thermicity associated with particle paths crossing the lead shock early in the cell sees a significant release of heat occurring around the ignition point, with a peak thermicity generally over 4-orders of magnitude greater than the pre-ignition thermicity. The peak thermicity appears as a very sharp increase in the highly unstable mixture, whereas this high heat release rate occurs over much longer times in weakly unstable mixtures. The peak thermicity is the most convincing method of tracking the ignition time of a particle, and this tracking method will be used for the remainder of this work. The peak thermicity attained along particle paths associated with quenched post-shock reactions is orders of magnitude lower than the peak thermicity attained along particle paths crossing the lead shock prior to the critical path. 

The evolution of thermicity allows for a clear differentiation of behaviour between two particle paths, but it alone cannot be used to determine the quenching along a particle path. This is especially true for ignition in argon-diluted hydrogen-oxygen, as the peak thermicity of the critical path ($\mathcal{O}(10^4)$) is similar to the following particle paths with quenched post-shock reactions ($\mathcal{O}(10^3)$). If one compares Figures \ref{fig:ThermicityEvolutionPlot} a) and b), despite a peak thermicity on the order of $10^4$ being associated with ignition in argon-diluted hydrogen-oxygen, a similar peak thermicity value is associated with quenching in methane-oxygen. This reinforces the need for another method for determining whether ignition occurs along a particle path. 

\begin{figure}
	\begin{subfigure}[c]{0.48\textwidth}
		\includegraphics[width=\linewidth]{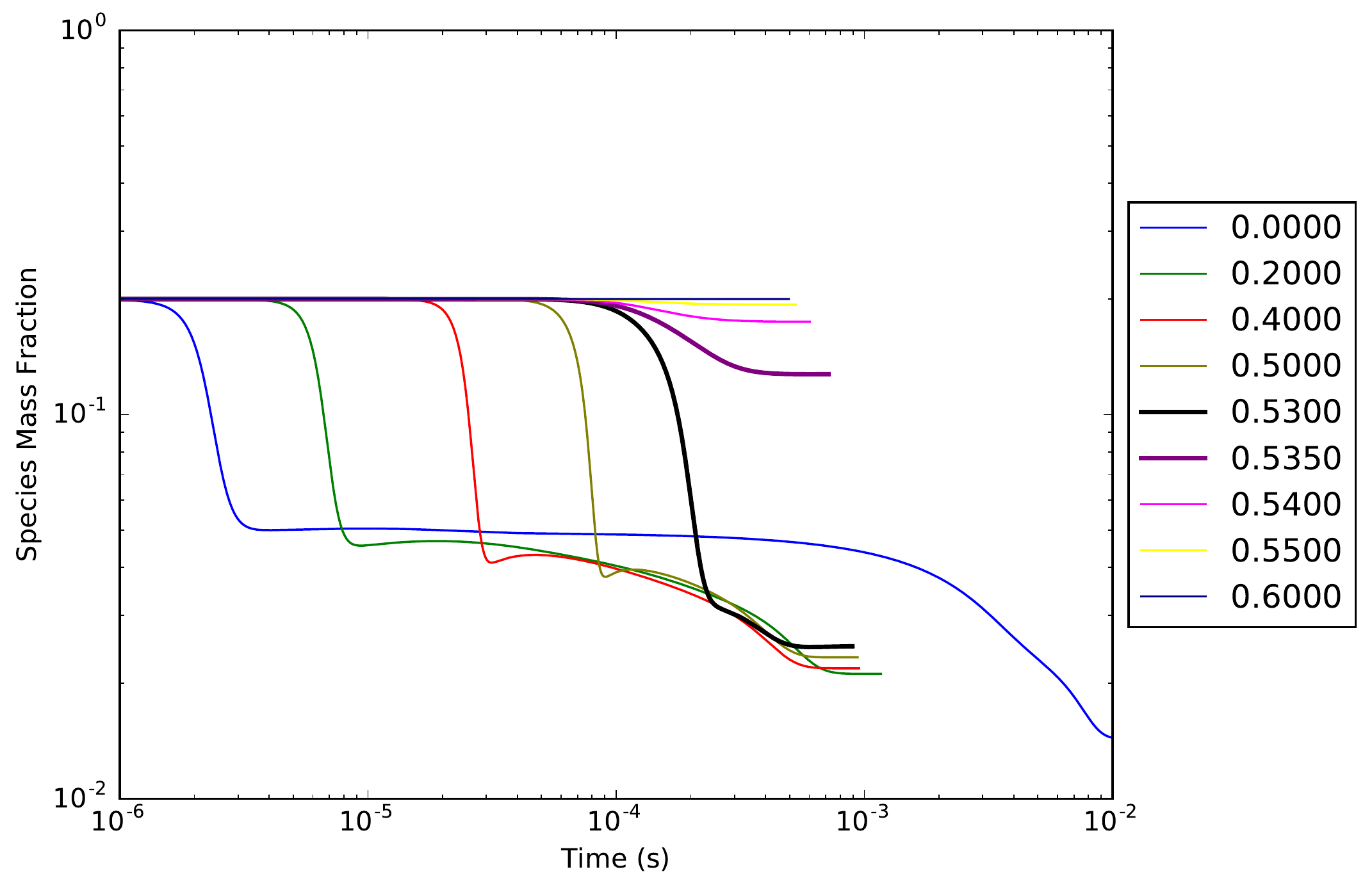} 
		\caption{}
	\end{subfigure}
	\begin{subfigure}[c]{0.48\textwidth}
		\includegraphics[width=\linewidth]{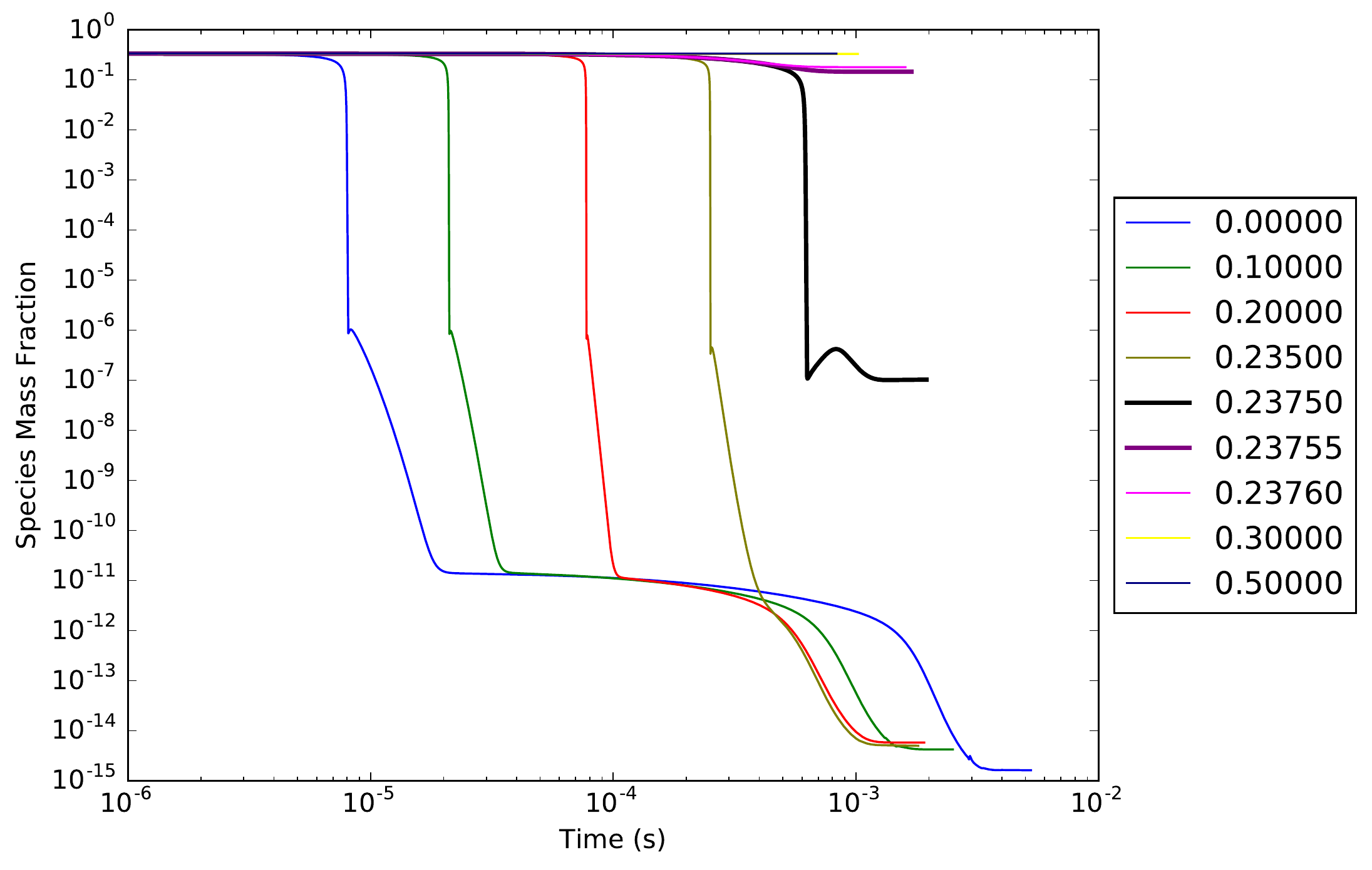} 
		\caption{}
	\end{subfigure}
	\caption{The evolution in time of the fuel species mass fraction (H$_2$ and CH$_4$), shown for selected particle paths. Plotted experiments are a) 13\_H2 and b) 0\_CH4, for which the critical ignition limit is associated with the particle paths 0.5300$L_{cell}$ and 0.2375$L_{cell}$.}
	\label{fig:ReactantsEvolutionPlot}
\end{figure}

A more convincing method to differentiate between particle paths associated with post-shock ignition and quenching appears when studying the evolution of the mass fractions of various chemical species as a function of time in the post-shock state.  Figure \ref{fig:ReactantsEvolutionPlot} shows the evolution of the mass fraction of the fuel specie, be it H$_2$ in a) or CH$_4$ in b), shown for the same particle paths as previously studied in this chapter. The difference between the particle paths which ignite and those which are quenched is very striking, with the reactant mass fraction suddenly decreasing during the ignition process. By contrast, particle paths associated with quenched reactions do not have this sudden drop in reactant mass fraction, with the final mass fraction of the reactants remaining comparable to the initial mass fraction of the reactants in the post-shock state. The critical path evolves similar to the earlier particle paths, reinforcing that ignition occurs along this selected path. Conversely, the following particle path (plotted in bold purple) with quenched post-shock reactions behaves as the subsequent particle paths reinforcing that no ignition occurs along this particle path. 

\begin{figure}
	\begin{subfigure}[c]{0.48\textwidth}
		\includegraphics[width=\linewidth]{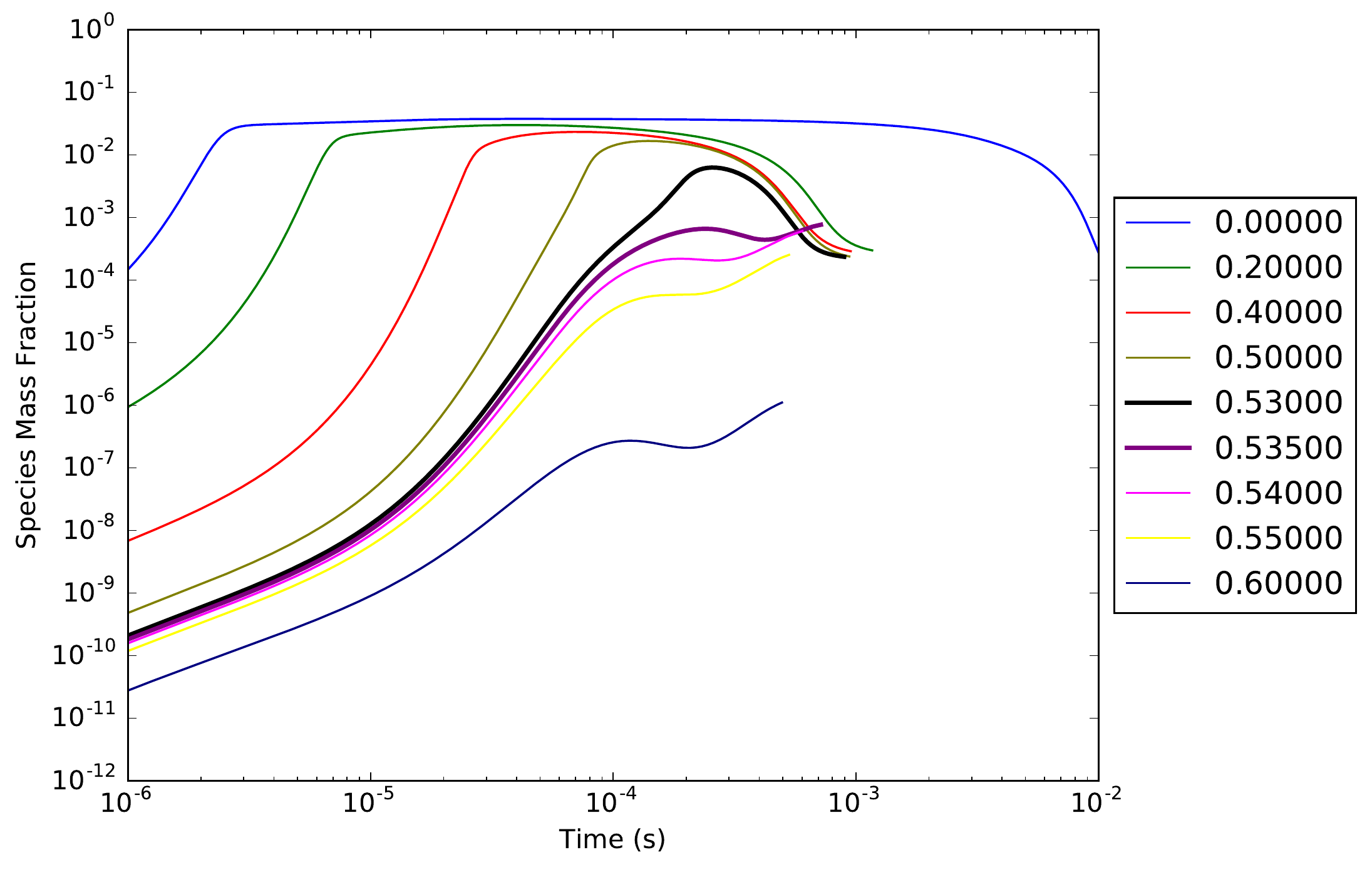} 
		\caption{}
	\end{subfigure}
	\begin{subfigure}[c]{0.48\textwidth}
		\includegraphics[width=\linewidth]{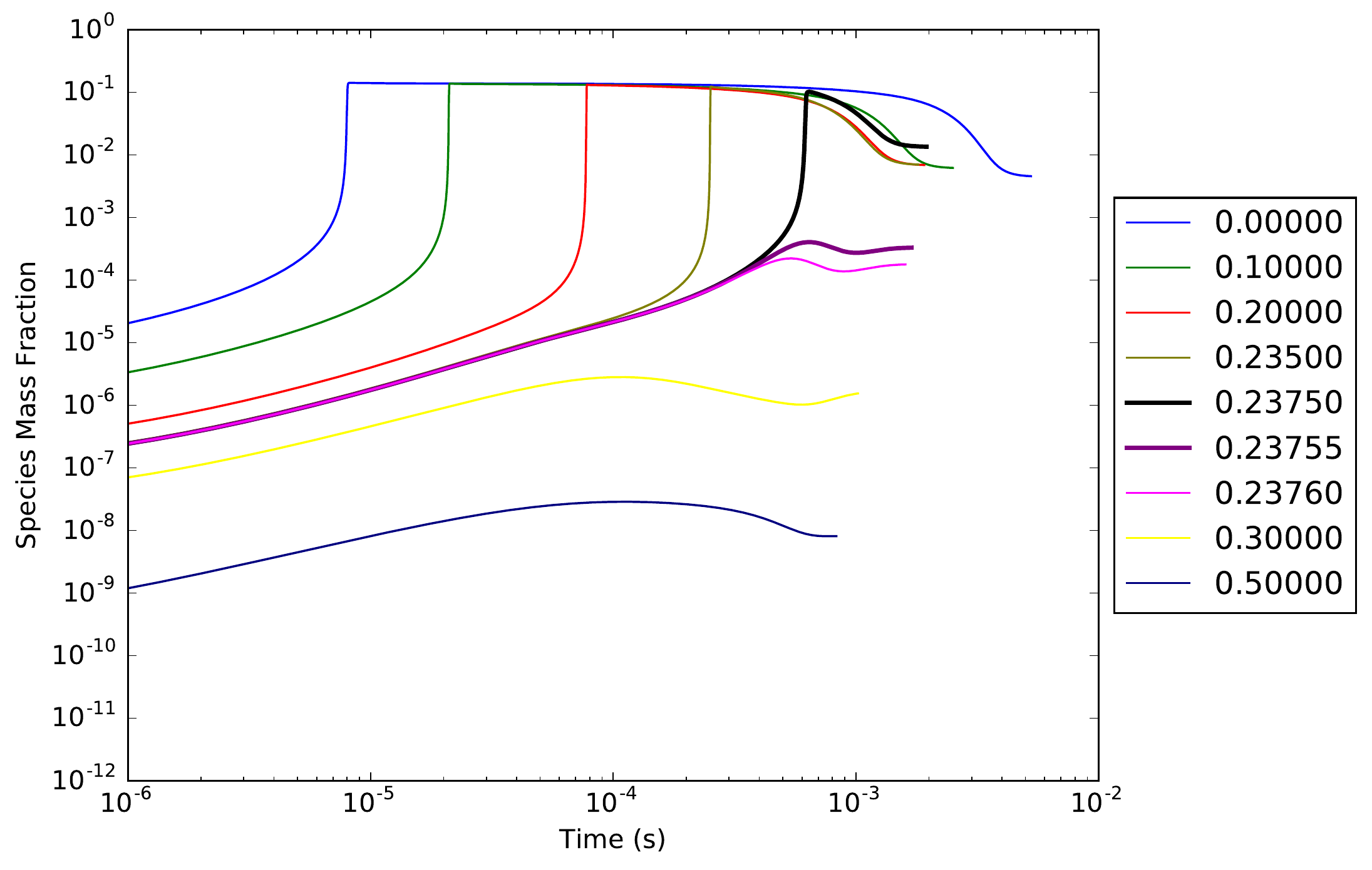} 
		\caption{}
	\end{subfigure}
	\caption{The evolution in time of an intermediate species mass fraction (OH), shown for selected particle paths. Plotted experiments are a) 13\_H2 and b) 0\_CH4, for which the critical ignition limit is associated with the particle paths 0.5300$L_{cell}$ and 0.2375$L_{cell}$.}
	\label{fig:IntermediatesEvolutionPlot}
\end{figure}

Intermediate species, such as H, O, and the radical OH, are very reactive and essential in the chain-branching reactions which can cause the overall chemical reaction to occur very rapidly. Intermediate species are consumed during chain-propagating and chain-termination elementary reactions. Previous studies have used the formation of OH radical as indicative of ignition.   A relevant example which is applicable to both currently-studied mixtures is the elementary reaction 
\begin{equation}
\mathrm{H} + \mathrm{OH} + X \longrightarrow \mathrm{H}_2\mathrm{O} + X,
\end{equation} 
in which two intermediate species are consumed to form a product specie in the presence of a third molecule $X$. The evolution of the mass fraction of the intermediate radical OH is shown in Figure \ref{fig:IntermediatesEvolutionPlot}. A noticeable increase in the OH mass fraction occurs during ignition. The particle paths associated with ignition in Figure \ref{fig:IntermediatesEvolutionPlot} a) have a peak mass fraction on the order of 10$^{-2}$ before decreasing near the end of the simulation. Meanwhile, the mass fraction of OH along particle paths with quenched post-shock reactions remain fairly insignificant, however they show that the radical is not consumed by chain-branching or chain-termination elementary reactions. A similar trend is observed in Figure \ref{fig:IntermediatesEvolutionPlot} b). Particle paths along which an ignition process occurs have a very sharp increase in the OH mass fraction, reaching an order of 10$^{-1}$ before being consumed later in the simulation. Conversely, the particle paths associated with quenching have insignificant OH mass fractions, which are not consumed by chain-branching or chain-termination elementary reactions.

The designation of a specific particle path as the critical path was not convincing when solely studying the evolution of temperature or thermicity in the post-shock state. In Figure \ref{fig:TemperatureEvolutionPlot} b), a rise in temperature still occurs for the first post-critical path associated with $x=0.23755 L_{cell}$, whereas the peak thermicity attained by the first post-critical path associated with $x=0.535 L_{cell}$ in Figure \ref{fig:ThermicityEvolutionPlot} a) is comparable to the peak thermicity attained along the critical path. However, when studying the evolution of the species mass fraction along these two paths, it becomes very clear that a Lagrangian particle propagating along the critical path ignites whereas a particle propagating along the first post-critical path does not. The occurrence of ignition along the critical path is supported by 1) the final mass fraction of the fuel species decreasing by an order of magnitude during ignition, 2) the mass fraction of the intermediate species increasing to a non-negligible level during the chain-initiation and chain-branching elementary reactions and subsequently decreasing during chain-termination elementary reactions which sees the formation of product species, and 3) the final mass fraction of the products of combustion exceeding the final mass fraction of the reactants. None of these statements hold true when studying a particle path which is linked to quenched post-shock reactions, despite certain similarities between the evolution of temperature or thermicity along these selected particle paths.

\subsection{Critical Ignition}
\begin{figure}
	\begin{subfigure}[c]{0.48\textwidth}
		\includegraphics[width=\linewidth]{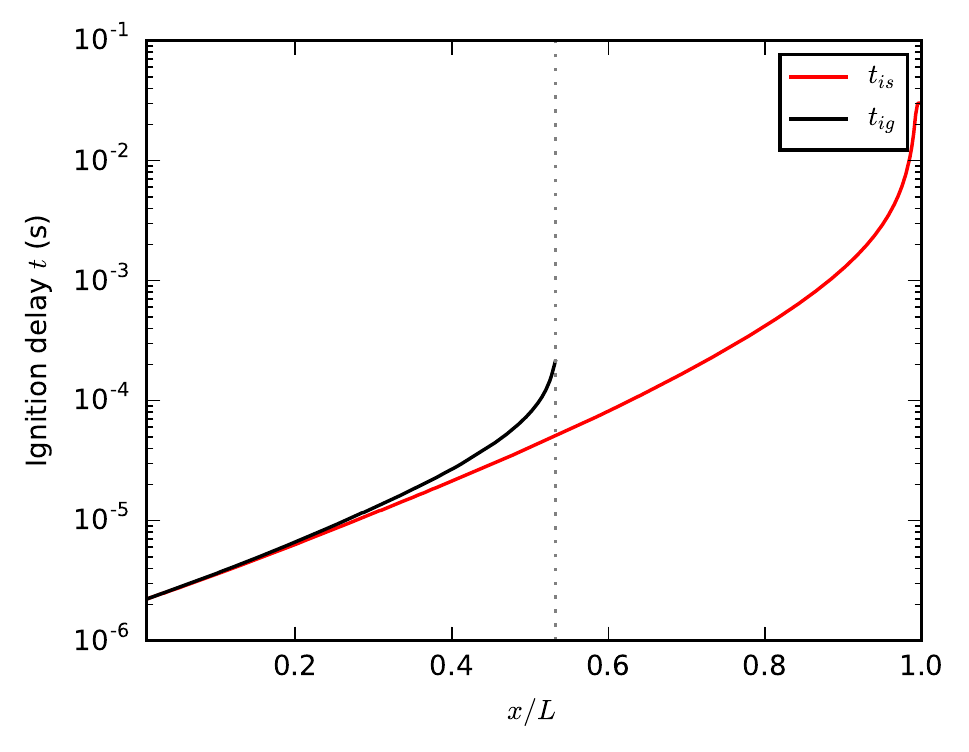} 
		\caption{}
	\end{subfigure}
	\begin{subfigure}[c]{0.48\textwidth}
		\includegraphics[width=\linewidth]{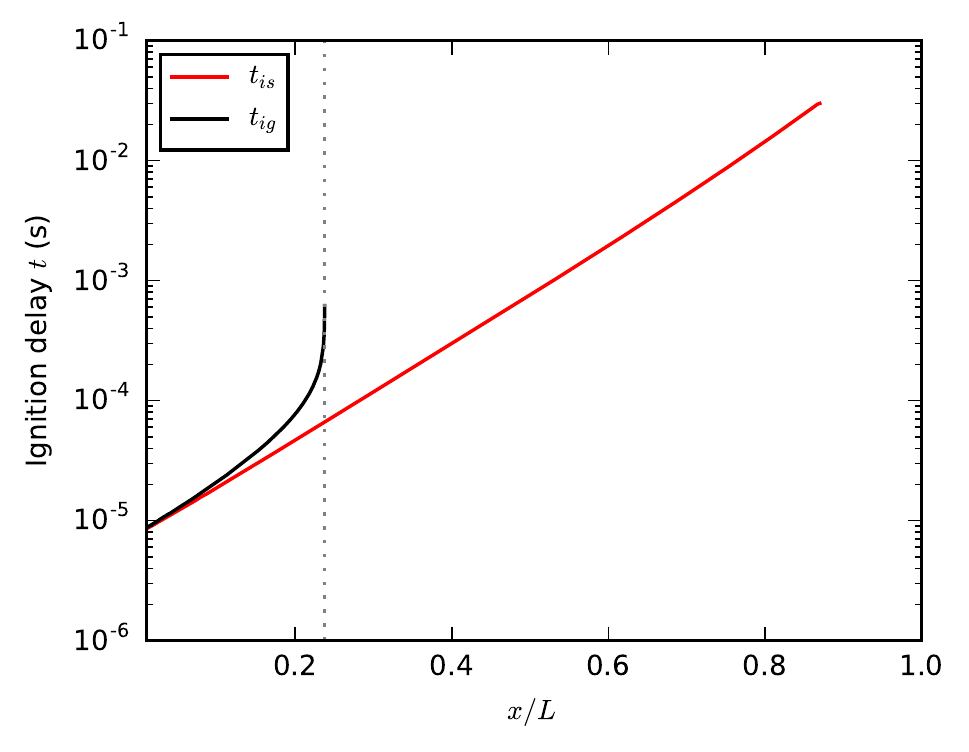} 
		\caption{}
	\end{subfigure}
	\caption{Evolution of the ignition delay over a detonation cell, with the critical ignition limit appearing as a vertical asymptote beyond which quenching of the post-shock ignition occurs. The effect of expansion cooling on the ignition delay is clearly seen when comparing the calculated ignition delays in the presence and absence of post-shock expansion. Plotted experiments are a) 13\_H2, and b) 0\_CH4.}
	\label{fig:CollapsedIgnitionDelay}
\end{figure}

Throughout the Cantera simulations, the peak rate of heat release is saved as a tracker for the ignition time. This remains the most convincing method of measuring the ignition delay, however studying the evolution of the species mass fraction remains necessary to distinguish between igniting and quenched particle paths. The ignition delay is plotted in Figure \ref{fig:CollapsedIgnitionDelay} over the length of a detonation cell, showing an ever-increasing ignition delay as the lead shock weakens. The ignition delay without expansion, $t_{is}$, is plotted alongside the ignition delay of a particle subjected to a volumetric expansion in the post-shock state, $t_{ig}$. The effect of expansion cooling is clearly seen as a vertical asymptote within the first half of the detonation cell. This asymptote represents the critical ignition limit of the cell, taken as the last particle path along which a Lagrangian particle will ignite. This clearly separates the cellular cycle into two distinct sections with the very different ignition behaviours seen in section \ref{ch:ChemKinetics}, be it an ignition regime prior to the critical ignition limit and a quenching regime after the limit. Physically, the compression of a particle caused by crossing the lead shock prior to this asymptote is strong enough to induce ignition, despite the expansion cooling affecting the particle throughout the post-shock region. Past this ignition asymptote, the lead shock has weakened and the volumetric expansion has increased to the point that the compression offered by the lead shock is unable to overcome the expansion cooling felt by a Lagrangian particle crossing the lead shock and induce auto-ignition. In experiments, secondary initiation mechanisms are necessary to ensure the ignition of these gases, such as the turbulent mixing of unburnt gases and heating by transverse shocks as seen in the experimental frames presented in section \ref{ch:Visualization}. The average critical ignition limit of the experiments performed in argon-diluted hydrogen-oxygen is situated around 53\% of the cell cycle and around 24\% of the cell cycle for the studied methane-oxygen experiment. This agrees with previous findings of quenching within the first half of the cell cycle. This also confirms that an accurate model for detonations must consider the lower heat release rate of secondary combustion mechanisms, as these mechanisms govern the energy release for the majority of the detonation cell. 

Numerical artefacts may appear when measuring the time to peak thermicity past the critical ignition limit, in the quenched reaction region. These are caused by the difficulties associated with capturing the peak thermicity during numerical integrations of the post-shock state for Lagrangian particles whose post-shock reactions are quenched. The time elapsed until the peak thermicity decreases immediately after the critical ignition limit is reached, and subsequent difficulties in capturing the peak thermicity appear once it becomes difficult to distinguish it from average thermicity. This is the case near the end of the cell, for particle paths affected by very large volumetric expansion rates. The particle paths crossing the lead shock immediately after the critical ignition limit were studied in detail, and the evolution of the species mass fraction confirmed no that ignition occurred along these particle paths. Despite these numerical artefacts, the time elapsed until reaching the peak thermicity remains a good measure of the ignition delay for particle paths prior to the quenching of post-shock reactions.

\subsection{Critical Ignition with Simplified Models}
\begin{figure}
	\begin{subfigure}[c]{0.48\textwidth}
		\includegraphics[width=\linewidth]{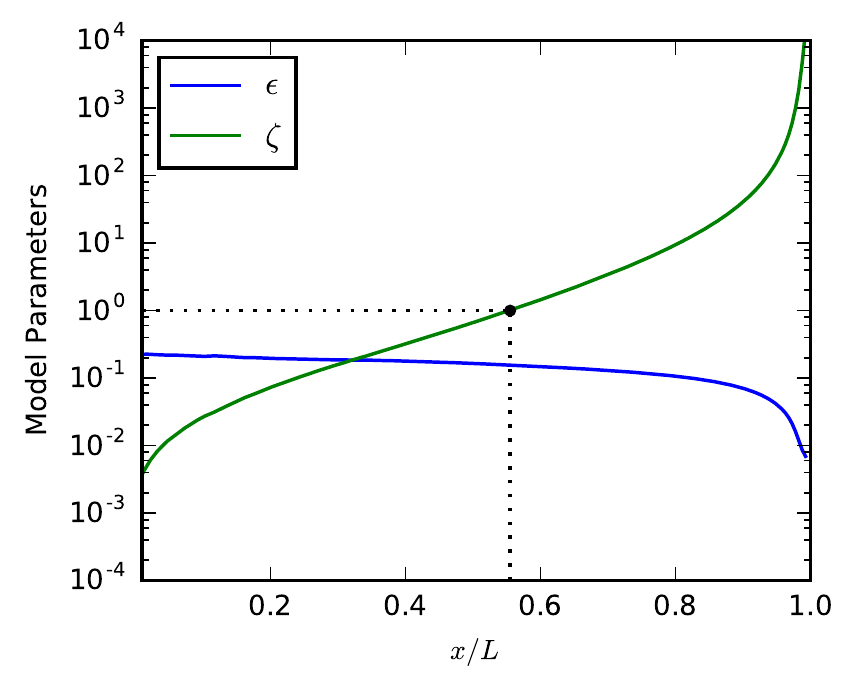} 
		\caption{}
	\end{subfigure}
	\begin{subfigure}[c]{0.48\textwidth}
		\includegraphics[width=\linewidth]{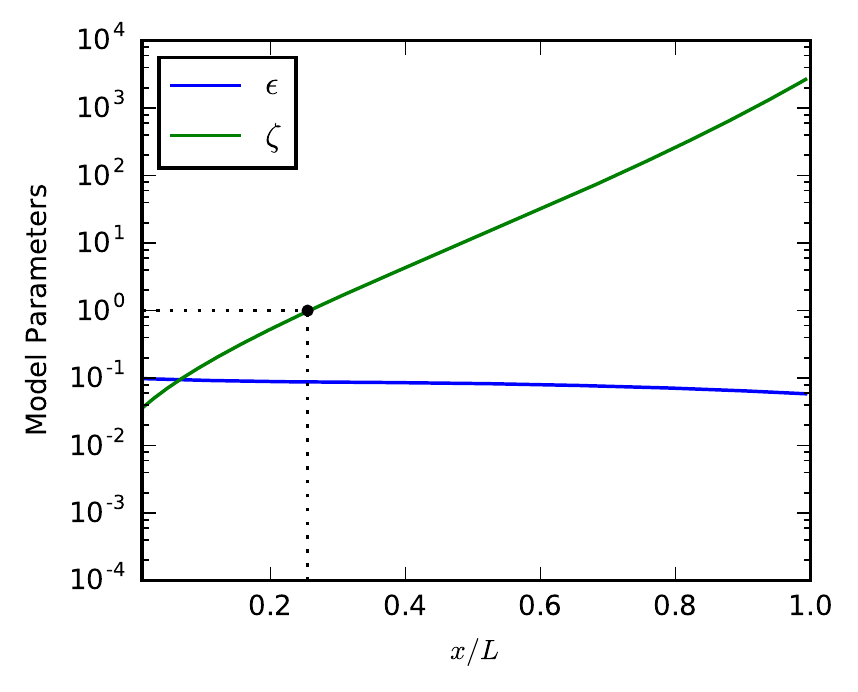} 
		\caption{}
	\end{subfigure}
	\caption{Evolution of the modelling parameters $\epsilon$ and $\zeta$ used in simplified combustion models plotted over a detonation cell. For experiments a) 13\_H2 and b) 0\_CH4. The dotted lines show the critical ignition asymptote predicted by the ignition criterion $\zeta<1$.}
	\label{fig:FittingParameters}
\end{figure}

Critical ignition can also be studied using simplified chemistry models to reduce the need for full chemistry calculations. The model parameters $E_a$ and $\zeta$ are plotted in Figure \ref{fig:FittingParameters} throughout the cell cycle. One can expect the ignition of a Lagrangian particle which crosses the lead shock at locations whose corresponding value of $\zeta$ satisfies the ignition criterion developed in section \ref{ch:IgnitionCriterion}, notably that $\zeta < 1$. The intercept for the critical ignition asymptote, located at $\zeta=1$ is plotted shown as a dotted line intercepting the x-axis at the predicted critical path. This path is located around $x/L=0.55$ and $x/L=0.25$ for experiments 13\_H2 and 0\_CH4, respectively. One immediately notices the excellent agreement between this prediction and the critical ignition limit shown in Figure \ref{fig:CollapsedIgnitionDelay}, calculated using a detailed chemical kinetics description of the post-shock reactions. As the modelling parameter monotonically increases throughout the cell, this criterion also models the quenching of post-shock reactions past the critical ignition limit, while allowing Lagrangian particles crossing the lead shock at distances prior to the critical ignition limit to ignite in finite time. For the majority of the cell cycle, representing a region of $0.2 < x/L < 0.8$, the controlling parameter and inverse activation energy can accurately be described by an exponential curve in space. 

\begin{figure}
	\begin{subfigure}[c]{0.48\textwidth}
		\includegraphics[width=\linewidth]{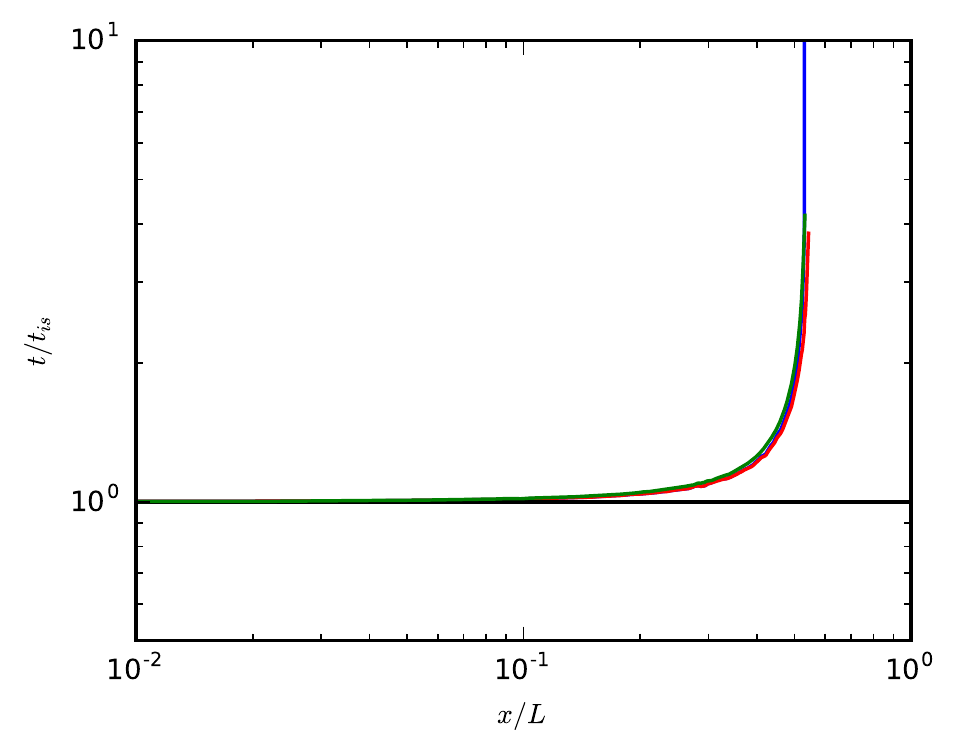} 
		\caption{}
	\end{subfigure}
	\begin{subfigure}[c]{0.48\textwidth}
		\includegraphics[width=\linewidth]{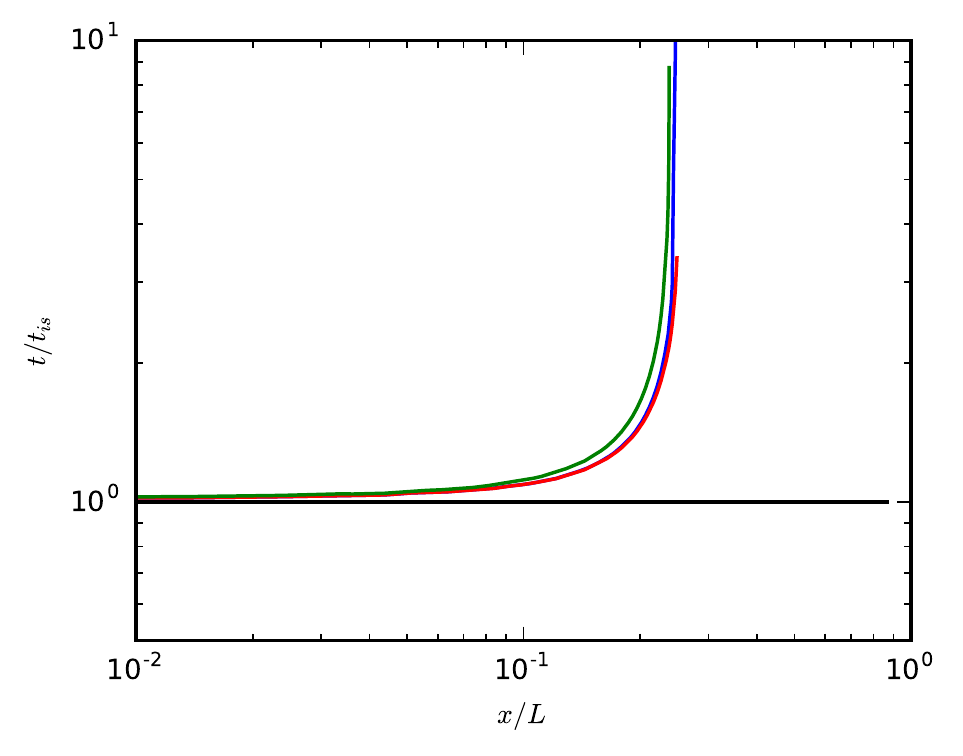} 
		\caption{}
	\end{subfigure}
	\caption{Comparison of the ignition delay evolution over a detonation cell calculated using a detailed chemistry model with those predicted by simplified models derived from 1-step and 2-step chemistry. For experiments a) 13\_H2 and b) 0\_CH4.  Blue : Implicit 2-step solution given in (\ref{eq:implicittin1}) for $n_i = \mathcal{O}(1)$. Red : 1-step and 2-step solutions in the limit of $\zeta = \mathcal{O}(1)$ given in (\ref{eq:ignitiondelay1step}) and (\ref{eq:ignitiondelay2stepapprox}). Green : solution obtained by integrating the post-shock state using a detailed chemical model. Black : ignition delay without expansion $t_{is}$.}
	\label{fig:ExperimentalIgnition}
\end{figure}

Using these calculated model parameters, one can meaningfully compare the ignition delay predicted by simplified chemical models and that calculated using a detailed chemistry model. Figure \ref{fig:ExperimentalIgnition} shows the evolution of the non-dimensionalized ignition delay over a detonation cell. One immediately notices that the implicit solution to the two-step model given by (\ref{eq:implicittin1}) best captures the ignition delay across both mixtures. This was also concluded in section \ref{ch:DetailedComparison} when comparing the ignition delay predicted by simplified models with that calculated using detailed chemistry in the context of ignition behind a shock propagating at its CJ velocity. The approximate 2-step model obtained by linearizing the implicit 2-step model in the limit of small $\epsilon$ captures the ignition delay the least well of any model, however the critical ignition asymptote is recovered within 3\% of that obtained using detailed chemistry. The final model obtained in the limit of $\zeta = \mathcal{O}(1)$ accurately predicts the ignition delay in the beginning of the cell, and recovers the critical ignition asymptote within the same 3\% as the approximate model. This not only supports the use of the critical ignition criterion developed in section \ref{ch:IgnitionCriterion} in the context of detonations, but also supports using simplified chemical models in the context of cellular detonations as an alternative to integrating a detailed chemical kinetics model. 

\section{Summary}
The development of a solution methodology to predict the ignition delay behind decaying shock waves, and the application of this theoretical treatment to assess the ignition mechanism in detonation cells.  

The ignition behind decaying shock waves was modelled in two parts.  The volume expansion rate along a particle path behind the shock was expressed in terms of the shock dynamic parameters (speed, acceleration and curvature) using the shock change equations.  This permitted to formulate 0D ignition problems along particle paths using realistic chemistry and simplified models using one or two steps.  The latter permitted to solve the problem in closed form and obtain the dependence of ignition delay on the shock dynamics and formulate an auto-ignition limit associated with a critical expansion rate.  

The results of the chemical modelling were then applied to gaseous detonations.   Shock tube experiments were performed with Schlieren visualization of low-pressure detonations propagating in the stable mixture of argon-diluted hydrogen-oxygen and the highly unstable mixture of methane oxygen.  The extreme differences in stability were chosen such as to span the entire range of known detonation stability in practice. From the photographs taken, the shock dynamics were determined as a function of the speed and curvature of the lead shock along the top wall, centreline, and bottom wall of the shock tube. These measurements allowed the reconstruction of an entire detonation cell. From the shock dynamics, the post-shock state was integrated using a detailed chemical kinetics model and the simple models developed in order to study the post-shock combustion and the quenching thereof inside the detonation cell.  

For both mixtures studied, ignition quenching was observed within the first half of the detonation cell.  Arguably, this suggests that all detonations in practice are characterized by this universal feature. The experiments showed however that the remainder of the gas crossing the lead shock after the critical ignition limit reacts from secondary initiation mechanisms such as turbulent mixing of unburnt pockets and combustion behind transverse shocks. The gas reacting due to secondary mechanisms comprises at least half the Lagrangian particles which cross the lead shock. This finding suggests that global gasdynamic models relying on lead shock auto-ignition alone are inadequate and further modelling of transverse shock compression and turbulent combustion is required.  Future work should extend this model to incorporate the compression of transverse shocks for modelling auto-ignition and turbulent burning of the remaining pockets of gas. 

Future work should also provide further validation of the proposed model for calculating the evolution of the ignition delay throughout a detonation cell, particularly the assumption of the constant characteristic volumetric expansion rate $ \left( -D( \ln \rho)/Dt \right)$ along a particle path.  This validation is a challenging task in experiments.  Although the experiments provide the instantaneous reacting fields, ignition is a convected phenomenon along particle paths, and the trajectory of particle paths is required in order to determine the time at which the particles crossed the shock.  This requires the detailed velocity field and its time evolution or models.  

This treatment of ignition behind a decaying shock links the information taken from the dynamics of the lead shock to the post-shock ignition dynamics. This was done using a local analysis at the lead shock which considered both the shock curvature and lead shock decay at the moment of crossing of a Lagrangian particle. This analysis gives the characteristic rate of expansion of that particle throughout the post-shock state, allowing for the integration of the conservation equations using this volumetric rate of change as a forcing function. This function is the link between the ignition dynamics and the dynamics of the lead shock which was assumed in previous works. 

The ignition delay predictions obtained using simplified models were validated against real gas calculations; very good agreement was found suggesting that both 1 step and 2 step models can be reliably used in reactive gasdynamics models characterized by large deformations possibly leading to ignition quenching. A unique criterion describing the occurrence or quenching of ignition behind the lead shock is presented, notably 
\begin{equation*}
\zeta = \frac{E_a}{RT_s}\left(\gamma-1\right) t_{is}(T_s) \left(\frac{-6}{\gamma+ 1}\frac{\dot{D}}{D} - 2D\kappa\frac{\gamma-1}{\left(\gamma+1\right)^2}\right) < 1.
\end{equation*}
This criterion generalizes the result of Eckett et al. and Radulescu \& Maxwell by accounting for the shock curvature. The experiments performed in a weakly unstable mixture of argon-diluted hydrogen-oxygen combined with the aforementioned analysis showed that there is decoupling of the lead shock with the post-shock reaction within the first half of the cell cycle. This is a novel finding for weakly unstable mixtures, as previous findings of such occurrences was limited to the highly unstable mixture of stoichiometric methane-oxygen.

\bibliographystyle{jfm}
\bibliography{jfm}

\end{document}